\documentclass{aa}  

\usepackage{graphicx}
\usepackage{txfonts}
\usepackage{subcaption}  
\usepackage{caption}
\usepackage{rotating} 
\usepackage{txfonts}  
\usepackage{hyperref}  
\usepackage{bm}
\usepackage{float}
\usepackage{url}
\usepackage{xcolor}
\usepackage{multirow} 
\usepackage{booktabs} 
\usepackage{siunitx}
\usepackage{threeparttable} 
\usepackage{natbib} 
\usepackage{tablefootnote} 
\usepackage{tabularx}
\usepackage{multicol}
\usepackage{dblfloatfix}      
\usepackage[section]{placeins}

\begin{document}

   \title{Comparing the data reduction pipelines of FRIPON, DFN, WMPL, and AMOS: Geminids Case Study}


   \author{P.M. Shober \inst{1}\fnmsep\thanks{corresponding authors}
          \and
          J. Vaubaillon \inst{1}
          \and
          S. Anghel \inst{1,2}\fnmsep$^{\star}$
          \and 
          H.A.R. Devillepoix \inst{3,4}
          \and 
          F. Hlobik \inst{5}
          \and 
          P. Matlovič \inst{5}
          \and 
          J. Tóth \inst{5}
          \and 
          D. Vida \inst{6}
          \and
          E.K. Sansom \inst{4}
          \and 
          T. Jansen-Sturgeon\inst{7}
          \and
          F. Colas \inst{1}
          \and 
          A. Malgoyre \inst{8}
          \and
          L. Kornoš \inst{5}
          \and 
          F. Ďuriš \inst{5}
          \and 
          V. Pazderová \inst{5}
          \and 
          S. Bouley \inst{9}
          \and
          B. Zanda \inst{10}
          \and
          P. Vernazza \inst{11}
          }

   \institute{LTE, Observatoire de Paris, Université PSL, Sorbonne Université, Université de Lille, LNE, CNRS, 61 Avenue de l’Observatoire, Paris, 75014, France 
         \and
             Astronomical Institute of the Romanian Academy, 5 - Cutitul de Argint Street, 040557 Bucharest, Romania
         \and 
             Space Science and Technology Centre, School of Earth
             and Planetary Sciences, Curtin University, Perth, WA 6845,
             Australia
         \and
             International Centre for Radio Astronomy Research, Curtin University, Perth, WA 6845, Australia
         \and 
             Faculty of Mathematics, Physics and Informatics, Comenius University, Bratislava, Slovakia
         \and 
             Department of Physics and Astronomy, University of Western Ontario, London, Ontario, Canada
         \and 
             STELaRLab, Lockheed Martin Australia, Adelaide, South Australia, Australia
         \and 
             Service Informatique Pythéas (SIP) CNRS – OSU Institut Pythéas – UMS 3470, Marseille, France
         \and 
             Université Paris-Saclay, UMR CNRS 8148, GEOPS, Orsay, France
         \and 
             Institut de Minéralogie, Physique des Matériaux et Cosmochimie, Muséum National d’Histoire Naturelle, CNRS, Paris, 75005, France
         \and
             Laboratoire d’Astrophysique de Marseille, Aix-Marseille University, CNRS, CNES, LAM, Institut Origines, 38 rue Frederic Joliot Curie, Marseille, 13388, France
             }

   \date{Received March 4, 2025; accepted October 20, 2025}
 
  \abstract
   {The number of meteor observation networks has expanded rapidly due to declining hardware costs, enabling both professional and amateur groups to contribute substantial datasets. However, accurate data reduction remains challenging, as variations in processing methodologies can significantly influence trajectory reconstructions and orbital interpretations.}
   {Our goal is to thoroughly compare four professionally produced meteor data reduction pipelines (FRIPON, DFN, WMPL, AMOS) by reprocessing FRIPON Geminid observations. Others can then use this analysis to compare their data reduction methods.}
   {We processed a dataset of 584 Geminid fireballs observed by FRIPON between 2016-2023. The single-station astrometric data is converted into the Global Fireball Exchange (GFE) standard format for uniform processing. We assess variations in trajectory, velocity, radiant, and orbital element calculations across the pipelines and compare them to previously published Geminid measurements.}
   {The radiant and velocity solutions provided by the four data reduction pipelines are all within the range of previously published values. However, there are some nuances. Particularly, the radiants estimated by WMPL, DFN, and AMOS are nearly identical. Whereas FRIPON reports a systematic shift in right ascension (-\,0.3$^{\circ}$), caused by improper handling of the precession. Additionally, the FRIPON data reduction pipeline also tends to overestimate the initial velocity (+0.3\,km\,s$^{-1}$) due to the deceleration model used as the velocity solver. The FRIPON velocity method relies on having a well-constrained deceleration profile; however, for the Geminids, many are low-deceleration events, leading to an overestimation of the initial velocity. On the other end of the spectrum, the DFN tends to predict lower velocities, particularly for poorly observed events. However, this velocity shift vanishes for the DFN when we only consider Geminids with at least three observations or more. The primary difference identified in the analysis concerns the velocity uncertainties. Despite all four pipelines achieving similar residuals between their trajectories and observations, their velocity uncertainties vary systematically, with WMPL outputting the smallest values, followed by AMOS, FRIPON, and DFN.}
   {From this Geminid case study, we find that the default FRIPON data reduction methods, while adequate for meteorite-dropping events, are not optimal for all cases. Specifically, FRIPON tends to overestimate velocities for low-deceleration events due to the less constrained fit, and the nominal radiants are not correctly output in J2000. Meanwhile, the other data reduction pipelines (DFN, WMPL, AMOS) produce consistent results, provided that the observational data is sufficiently robust, i.e., more than $\sim$50 data points from at least three observers. A key takeaway is the need to re-evaluate how velocity uncertainties are estimated. Our results show that uncertainty estimates vary systematically across different pipelines, despite generally similar goodness-of-fit statistics. The increasing availability of impact observations from varying sources (radar, video, photo, seismic, infrasound, satellite, telescopic, etc.) calls for greater collaboration and transparency in data reduction practices.}


   \maketitle

\section{Introduction}
Fireball observations have historically relied on film-based techniques \citep{ceplecha1961multiple,mccrosky1971lost,halliday1978innisfree}, but the advent of charge-coupled devices (CCDs) and, more recently, complementary metal–oxide–semiconductor (CMOS) sensors has revolutionised both the scale and precision of meteor and fireball surveys \citep{molau1999meteor,koten2019meteors,IMO_Handbook_2022}. 

Photographic observations generally excel in spatial resolution and yield very precise astrometric solutions, making them particularly well-suited for recovering meteorites. However, because photographic techniques rely on relatively few discrete exposures (e.g., one exposure spanning minutes to hours), they provide limited time resolution; abrupt changes in meteor brightness, detailed velocity variations, or short-duration fragmentation episodes can be difficult to capture. By contrast, video observations utilise continuous, frame-based recordings, typically at 25–60 frames per second (or higher), providing near-real-time detection and automated data processing. This frame-by-frame approach provides excellent temporal resolution at the cost of lower spatial resolution. However, this shortcoming can be mitigated by employing high-definition sensors or deploying multiple narrow-field cameras to enhance positional accuracy. As commercial CCDs and later CMOS sensors matured, sensitive yet affordable camera hardware proliferated in both professional and amateur networks \citep{jeanne2019calibration,koten2019meteors,IMO_Handbook_2022}.

Film-based photographic observatories continued into the 21st century for bright fireball observations, but as digital sensors shrank in size, improved in sensitivity, and fell in cost, analogue film gave way to digital imaging \citep{spurny2006automation,howie2017build}. This digital revolution has spawned large-scale photographic fireball observation networks that were previously infeasible due to geography or expense \citep{bland2012australian,koten2019meteors,devillepoix2020global}, leading to a significant increase in data collection and meteorite recoveries. In the four decades between 1959 and 1999, only five meteorite falls (P\v{r}ibram, Lost City, Innisfree, Bene\v{s}ov, Peekskill) were recovered based on instrumental observations. By comparison, in 2023 alone, meteorites from three distinct falls were recovered on three consecutive days: 13 February in France \citep{zanda2023rrecovery}, 14 February in Italy \citep{barghini2024recovery}, and 15 February in Texas\footnote{\url{https://ares.jsc.nasa.gov/meteorite-falls/events/el-sauz-tx}}. 

A diverse array of meteor and fireball networks now exists worldwide, ranging from higher-precision, photographic systems like the DFN \citep{howie2017build,devillepoix2020global} and the EN \citep{borovivcka2022_one}, to more modular video approaches like FRIPON \citep{colas2020fripon} and the Global Meteor Network \citep{vida2021global}. Additional examples include the Spanish Meteor Network \citep{trigo2006spanish}, NASA’s All-Sky Fireball Network \citep{cooke2012status}, the Cameras for all-sky Meteor Surveillance (CAMS) project \citep{jenniskens2011cams}, SonotaCo’s collaborative effort in Japan \citep{sonotaco2009meteor}, Polish Fireball Network \citep{wisniewski2017current}, and the AMOS initiative in Slovakia \citep{toth2015all}, as well as a growing number of amateur-run networks (e.g., all-sky7; \citealp{hankey2020all}). Although some focus on near-real-time, frame-based video while others opt for longer-exposure, higher-resolution imaging, many of the calibration and data-reduction methods are effectively the same \citep{koten2019meteors}. Each network’s hardware choices are shaped by its scientific goals and expense limitations.

Building reliable links between observable parameters (e.g., light curves and velocity profiles) and the internal processes (e.g., ablation, mass loss, fragmentation) requires careful theoretical modelling. A meteoroid’s flight through Earth’s atmosphere involves multiple interacting processes -- shock wave generation, aerodynamic heating, ablation and mass loss, fragmentation, ionisation, and the resulting luminous emission -- each influencing the meteor’s brightness, trajectory, and outcome. Additionally, the unknown parameters of the meteoroids, such as shape, spin, mass, and composition, further complicate this issue. Early models often adopted ``single-body'' ablation, in which the mass of a compact object with cross-section $S$ evolves under standard drag and ablation equations \citep{Bronshten1983, ceplecha1998meteor}. This approach remains valuable in many automated or large-scale fireball networks, where a linear or straight-line fit to the geometric path is common \citep[e.g.,][]{Borovicka_1990BAICz, gural2012new}. Such simplifications can be sufficient under favourable conditions or for specific events. However, recent work has underscored that fragmentation is pervasive and can drastically alter a meteoroid’s velocity profile and final mass \citep{ceplecha1993atmospheric,borovicka2003moravka,borovivcka2020two,borovivcka2022_one,borovivcka2022_two}, necessitating more robust dynamical models and careful uncertainty quantification. A primary challenge, as emphasised by \citet{egal2017challenge}, lies in accurately deriving the velocity from photographic or video records. Many early methods that separate the geometric and velocity fits -- such as the straight-line least-squares approach \citep{Borovicka_1990BAICz} -- may underestimate uncertainties when the observations have complex geometry or when the meteoroid undergoes multiple fragmentation episodes. In particular, \citet{egal2017challenge} highlights that even small systematic biases in frame timing or astrometric precision can lead to errors in the derived deceleration profile, with direct consequences for orbit determination. This issue becomes especially critical in low-convergence-angle scenarios or when the event has a protracted, curved path.
Advanced pipelines increasingly employ dynamic trajectory solvers that integrate the meteoroid’s three-dimensional equations of motion directly into the data-fitting step \citep{gural2012new,vida2020estimating, jansen2020dynamic}. Rather than first inferring a simple geometric path and then fitting a velocity separately, these approaches can simultaneously solve for position and velocity at each timestep, often incorporating Monte Carlo or particle-filter methods to map uncertainties more fully \citep{jansen2020dynamic, sansom20193d}. 

Along similar lines, \citet{egal2017challenge} stresses that reliably computing velocity requires careful propagation of all measurement errors -- most critically timing, pointing astrometry, and any potential system offsets -- to avoid unrealistically low or inconsistent uncertainties.
These modern methods reflect a broader push toward Bayesian or stochastic approaches that can capture the inherent complexity of meteor flight and assign more realistic error bars. While most authors now agree that velocity uncertainties below $\sim0.1$\,km\,s$^{-1}$ are necessary to reliably discriminate orbital source regions \citep{granvik2018debiased}, actual uncertainties in automated or semi-automated networks sometimes exceed this figure due to sensor calibration, partial fragmentation modelling, or constraints in processing time. Many networks still rely on simplified single-body or straight-line fits in their automated pipelines, principally for operational efficiency. 

We are collecting more and more data on the meteoroids that continuously impact our atmosphere \citep{koten2019meteors}. It is excellent to have increasingly large meteor and fireball datasets. Albeit, we must be cautious. Despite the rapid amelioration of hardware costs, enabling the development of many more professional networks and even amateur networks, analysing the collected impact data is still complicated and difficult to reduce. We are using this data to better understand the meteoroid environment in near-Earth space, the source regions of meteorites, and the physical and orbital evolution of small-body populations. Thus, we must be transparent about our data-reduction methods, as slight variations or systematic errors in data reduction can significantly alter how we interpret the results and their implications for our understanding of small bodies in the solar system. This is especially important as many networks, particularly video-based, are increasingly fully automated in their data acquisition, processing, and reduction \citep{molau2001akm,gural2009new,brown2010development,gural2011california,weryk2013canadian,suk2017automated,koten2019meteors,colas2020fripon,pena2021accurate,vida2021global}. 

In today's modern research environment, we must be increasingly adept at sharing our code so that others may test and scrutinise it. The first clear example was the observation and recovery of the Winchcombe meteorite fall in the UK in February 2021. This carbonaceous meteorite fall was observed by five different camera systems and six networks in total. Due to the extraordinary nature of the fall and observations, it was necessary for the first time to combine and coordinate the observations and pipelines of several fireball networks. This collaboration led to the development of the Global Fireball Exchange (GFE) format, a universal method of sharing the detailed single-station observations of a fireball event \citep{rowe2021just}. Networks then implemented this format so that they could all communicate and process the data in coordination, leading to the recovery of the meteorite and its complete characterisation \citep{King_Winchcombe2022SciA,mcmullan2023winchcombe}.

This work aims to assess these cross-network variations and promote best practices for ensuring consistent, reliable results across different pipelines. We compare four established, professionally developed fireball data-reduction pipelines, all tested against actual FRIPON network observations. Our cross-evaluation highlights differences in calibration strategies, trajectory modelling approaches, and orbital solutions, offering insights for more convergent standards in future multi-network meteor research.

The paper is organised as follows: Section \ref{sec:pipelines} introduces the four contributing networks and their pipelines. Section \ref{sec:method} outlines the data sets and processing methods, while Section \ref{sec:results} compares the derived trajectories and orbital elements. We discuss implications for multi-network validation in Section \ref{sec:discuss}, and finally, Section \ref{sec:howto} explores strategies for improving meteor data reduction and future collaborative efforts.

\section{Networks and software used} \label{sec:pipelines}
\subsection{FRIPON}\label{sec:FRIPON}
Initiated in 2015, FRIPON is a pioneering French-led project dedicated to the detailed observation of fireballs and the recovery of meteorites. This collaborative effort now spans 15 countries on four continents, incorporating over 220 cameras covering approximately 2 million km~$^{2}$. Using all-sky CCD cameras that record at 30 frames per second, FRIPON achieves high-time-resolution data capture. The network has a detection limit of about zero magnitude, making it tuned for fireball observation and meteorite recovery. Extended exposures are taken every 10 minutes to enhance astrometry and photometry, achieving effective signal-to-noise ratios up to magnitude 6 in optimal conditions\citep{anghel2019photometric,jeanne2019calibration,jeanne2020méthode,colas2020fripon}. The automated system of FRIPON enables swift mobilisation for meteorite recovery efforts, focusing on objects estimated to weigh 500 grams or more, by modelling the impact via photometric and dynamic methods \citep{2021MNRAS.508.5716A, jeanne2019calibration}. To date, the global expansion of the consortium has aided in the recovery of five meteorites, marking significant strides in the field of meteorite retrieval and interplanetary study \citep{gardiol2021cavezzo,mcmullan2023winchcombe,Antier_Par,barghini2024recovery,Egal2025_SPLV}. The data reduction process within FRIPON is described in detail in \citet{jeanne2019calibration}, \citet{jeanne2020méthode}, and \citet{colas2020fripon}. The processed data collected by the FRIPON project is publicly available\footnote{\url{https://fireball.fripon.org/}}.

FRIPON estimates the internal error of each camera ($\overline{\sigma_i}\simeq$~0.75~\text{arcmin}; \citealp{colas2020fripon}) by fitting a reference plane through all picks. A local comparison (100~px box) between measured and HIPPARCOS positions gives the systematic term $s_i$ \citep{bessell2000hipparcos}. The straight–line trajectory $\mathcal T$ is obtained by minimising  

\begin{equation}
  S(\mathcal T)=
  \sum_{i=1}^{n_{\mathrm{cam}}}
  \sum_{j=1}^{n_i}
        \frac{\epsilon_{ij}(\mathcal T)^2}
             {\sigma_i^{2}+n_i s_i^{2}},
  \label{eq:friponchi2}
\end{equation}

\noindent
where $\epsilon_{ij}(\mathcal{T})$ is the residual between the $j$-th measure taken by the $i$-th camera and the candidate trajectory $\mathcal{T}$, $s_{i}$ is its systematic error, $\sigma_{i}$ is the internal random error of the $i$-th camera, and $n_{i}$ is the number of images acquired by that camera. FRIPON uses this method to characterise the systematic errors of our cameras (e.g., a misaligned lens); however, this approach doesn't address errors such as the camera's location. To address these errors, FRIPON computes an initial estimate of the trajectory and compares the residuals with the expected random and systematic errors. If they are larger than expected for a specific camera, they iteratively decrease its weight during the calculation of the trajectory. The final systematic error is usually on the order of 0.3 arcmin, which ends the iterative process.

The entry speed is recovered based on the single–body drag/ablation model of \citet{Bronshten1983} and \citet{stulov1995aerodynamics}. 
Methods such as \citet{gritsevich2009determination} reformulate these without assuming any parameters, enabling a dimensionless coefficient approach. The FRIPON pipeline employs the method described by \citet{jeanne2019calibration}. Although similar to \citet{gritsevich2009determination}, it uses a different set of equations that do not make use of the dimensionless parameters:

\begin{equation}
\begin{aligned}
\frac{\mathrm d V}{\mathrm d t} &=
  -\tfrac12\,A\,\rho_{\mathrm{atm}}\,V^{2}\,
  \exp\!\Bigl[\tfrac{B(1-\mu)}{2A}({V^{2}-V_e^{2}})\Bigr],\\[4pt]
m &=
  \exp\!\Bigl[\tfrac{B}{2A}(V^{2}-V_e^{2})\Bigr].
\end{aligned}
\label{eq:AB}
\end{equation}

\noindent
with a deceleration parameter 
$A=c_dS_e/M_e$ and an ablation parameter $B=c_hS_e/(HM_e)$, where $c_d$ is the drag coefficient, $c_h$ the heat–transfer coefficient, $H$ the enthalpy of destruction, $\rho_{\mathrm{atm}}$ the atmospheric gas density (NRLMSISE-00 model; \citealp{lyytinen2016implications}), $m$ the normalised meteoroid mass, $M_e$ the pre-entry mass, $s$ the normalised cross-sectional area, $S_e$ the pre-entry cross-sectional area, and $\mu$ the shape-change coefficient.
Fitting the projected picks on the straight line \citep{jeanne2019calibration} yields the parameters $V_e$, $A$, and $B$ and their joint confidence region. The estimated $V_{e}$ value represents the velocity at the top of the atmosphere, within FRIPON this is arbitrarily defined as 10\,km above the first measured point. 

\begin{table*}[tbp]
    \caption{Comparison of trajectory, velocity solvers and outlier removal approaches.}
    \label{tab:velocity_comparison}
    \centering
    \begin{threeparttable} 
    \begin{tabularx}{\textwidth}{|l|p{0.12\textwidth}|X|X|X|p{0.06\textwidth}|}
    \hline \hline
    \textbf{Pipeline} & \textbf{Designed For} & \textbf{Trajectory Solver} & \textbf{Velocity Solver} & \textbf{Outlier Removal?} & \textbf{Sources} \\ \hline
    FRIPON   & Lower-resolution video, fireballs  & Straight-line least squares \citep{Borovicka_1990BAICz} & Fits physically based single-body differential equations that model meteor deceleration and ablation, rewritten into two independent equations by \citet{jeanne2019calibration} following the method of \citet{turchak2014meteoroids,gritsevich2009determination}. & Yes & † \\ \hline
    DFN      & Higher-resolution photo, fireballs & Straight-line least squares \citep{Borovicka_1990BAICz} & Meteoroid initial states ($M_e$ and $V_e$) are estimated using an Extended Kalman Filter on straight line triangulated positions (measurements). This updates estimates of position, mass, and velocity throughout the trajectory, while modelling uncertainties. & No & ‡ \\ \hline
    WMPL     & Higher-resolution video, meteors & Monte Carlo trajectory solver: initial geometry via IP/LoS, followed by noise injection (based on measured angular residuals) and global minimisation of a timing cost function & A linear “lag” fit is performed on the early 25–80\% of the meteor’s path (where deceleration is minimal). The best fit, i.e., the one with the smallest standard deviation, provides the initial velocity estimate. & Outlier weighting is applied on a per-measurement basis (with lower weights for observations with poor perspective) & § \\ \hline
    AMOS     & Lower-resolution video, meteors, spectroscopy  & Intersecting planes and Straight-line least squares methods \citep{1987BAICz..38..222C,Borovicka_1990BAICz} & Exponential fit of decelerated meteors along the length of trajectory, otherwise linear fit. & Yes & ¶ \\ \hline
    \end{tabularx}
    \tablefoot{``Higher-resolution'' implies pixel scale $\sim$1 arcminute; ``lower-resolution'' implies pixel scale $\sim$10 arcminutes.}
    
    \begin{tablenotes}
        \footnotesize
        \item[†] \citet{jeanne2019calibration}, \citet{jeanne2020méthode}, \citet{colas2020fripon}
        \item[‡] \citet{howie2017build}, \citet{sansom2015novel}, \citet{devillepoix2020global}
        \item[§] \citet{vida2018modelling}, \citet{vida2020estimating}
        \item[¶] \citet{Duris2018}, Tóth (in prep.)
    \end{tablenotes}
    
    \end{threeparttable}
\end{table*}

\subsection{DFN}\label{sec:DFN}
The Desert Fireball Network (DFN), established to cover the expansive Australian outback, surveys over 2.5 million km$^{2}$. This equates to over a third of the Australian continent, underscoring the DFN's extensive observational reach since its digital transition in 2013-2015 \citep{bland2012australian,howie2017build}. The deployment of 50 digital fireball observatories, equipped with high-resolution DSLR cameras and all-sky fisheye lenses, enables the network to achieve unprecedented capture rates of meteoroids, thereby enhancing the precision of meteorite fall position determinations through GNSS-synchronised liquid crystal shutters \citep{howie2017submillisecond,devillepoix2019observation}. The precision and manual picking of fireball observations have resulted in a vast, accurate fireball dataset that helps us better understand the dynamics of centimeter- to meter-sized debris in the inner solar system \citep{devillepoix2019observation,sansom2019determining,shober2020did,shober2021main,shober2025perihelion}. The DFN's automated methodology for determining atmospheric trajectories and extracting velocity profiles utilizes a refined straight-line least squares approach coupled with an extended Kalman smoother, ensuring accurate trajectory and velocity data while accounting for observational and fitting uncertainties \citep{Borovicka_1990BAICz,sansom2015novel}.

Expanding beyond its Australian origins, the DFN now forms the core of the Global Fireball Observatory (GFO)\footnote{\url{https://gfo.rocks/}}, integrating ten partner networks and 18 collaborating institutions across nine countries, all employing DFN-developed observatories \citep{devillepoix2020global}. The GFO has led to the retrieval of 18 orbital meteorites in total\footnote{\url{https://dfn.gfo.rocks/meteorites.html}}, representing about 30\% of all meteorites recovered with known orbits to date \citep{bland2009anomalous,dyl2016characterization,devillepoix2018dingle,jenniskens2019creston,sansom2020murrili,King_Winchcombe2022SciA,shober2022arpu,Devillepoix_Madura_Cave,anderson2022successful,brown2023golden}. 

\subsection{AMOS}\label{sec:AMOS}
Based in Slovakia, the All-Sky Meteor Orbit System (AMOS)\footnote{\url{https://amos.uniba.sk}} is an advanced global meteor observation network that utilises automated intensified all-sky video cameras to track and analyse meteor events, and wide-field spectral cameras to study meteoroid composition. Established in 2009 by the Faculty of Mathematics, Physics, and Informatics at Comenius University Bratislava, AMOS aims to provide coverage over Central Europe with a network currently spanning six stations in Slovakia and, in total, 17 stations around the globe. Other network locations include the Canary Islands, Chile, Hawaii, Australia, and South Africa. Each station has a digital CCD camera capable of recording high-resolution video (1600x1200) at up to 20 frames per second, covering a wide range of meteor magnitudes, typically between +4.0 and -3.0 in stellar magnitude, including fireballs as well \cite{toth2015all}. AMOS's primary goal is to calculate precise trajectories and orbits of meteors and correlate them with their spectral properties studied by the AMOS-Spec cameras \citep{matlovic2020} to improve the knowledge of weak meteor showers and their parent body characterisation, or to provide comprehensive studies of meteoroid dynamics, physical properties and composition \citep{matlovic2019, 2022MNRAS.513.3982M}. The network has successfully integrated cutting-edge software algorithms to determine the atmospheric paths, radiant points, and orbital parameters of meteors, utilising methods like the Least Squares Monte Carlo fitting technique \citep{Duris2018} for enhanced accuracy in trajectory reconstruction \citep{toth2019}. These cameras are part of an integrated system designed for advanced meteor detection and analysis, contributing to research on meteor trajectories, origins, and potential recovery of meteorites.

\subsection{WMPL and RMS}\label{sec:RMS}
The Western Meteor Python Library (WMPL), developed by the Meteor Physics Group at the University of Western Ontario in Canada, is a comprehensive open source\footnote{WMPL library on GitHub: \url{https://github.com/wmpg/WesternMeteorPyLib} (accessed July 10, 2024)} toolset written in Python designed to process, analyze, and interpret the observational data obtained from various meteor detection systems, facilitating detailed analysis of meteor trajectories. The library includes a variety of commonly used methods in the meteor field \citep{vida2018modelling}, a novel meteor trajectory solver \citep{vida2020estimating}, a novel method for meteor shower mass index estimation \citep{vida2020new}, and implementation of the \cite{borovivcka2007atmospheric} faint meteor model \citep{vida2024first}, and implementation of the \cite{borovivcka2013kovsice} semi-empirical fireball fragmentation model \citep{vida2023direct}. 

Applicable to this work, WMPL's meteor trajectory solver uses both the geometric and dynamical data to constrain meteor trajectories and applies rigorous statistical methods, including Monte Carlo simulations, to fully propagate the uncertainties in all observed trajectory parameters and derived orbital elements. The transparent nature of WMPL represents a significant step towards standardising the methodology used in the field. WMPL has been adopted as the main trajectory analysis toolset for several major meteor networks, including the originating Global Meteor Network \citep{vida2021global}, all-sky7 \citep{hankey2020all}, the UK Meteor Network \citep{campbell2014uk}, and the NASA fireball network \citep{cooke2012status}.

In addition to WMPL, the Raspberry Pi Meteor Station (RMS) library\footnote{RMS library on GitHub: \url{https://github.com/CroatianMeteorNetwork/RMS} (accessed July 10, 2024)} provides universal tools for reducing optical meteor data (both photographic and video), implementing the latest astrometric and photometric calibration methods \citep{vida2021global}. The measurements made by RMS tools are fully compatible with the international standards \citep{rowe2021just} and can be transparently used in WMPL.

\section{Methods}\label{sec:method}

\subsection{FRIPON Observations}
When integrating and comparing meteor data across different networks, it is crucial to account for variations in hardware, observing cadence, exposure lengths, network density, and calibration and reduction methods. These factors can significantly influence the accuracy of the trajectory and origin estimations and the overall interpretation of the meteor data. These factors -- ranging from sensor resolution and frame rate to the specific observational strategies employed -- directly influence the accuracy and precision of derived trajectories, orbital elements, and their corresponding uncertainties.

The FRIPON observatories use a CCD Sony ICX445 chip with 1296 × 964 pixels and a pixel size of 3.75 × 3.75 µm. The lens is a fish-eye with a 1.25 mm focal length opened to f/2, giving a pixel scale of 10 arcmin \citep{colas2020fripon}.
These observatories are lower resolution than those used by the EN or DFN ($\sim$1-2\,arcmin, i.e., 5-10$\times$ the resolution), which both utilise automated fireball observatories with high-resolution DSLR cameras  \citep{howie2017build,borovivcka2022_one}. Despite the lower optical resolution of the FRIPON sensors, the network operates and aims to recover meteorites through its higher network density and increased frame rate. The CCD observations made by FRIPON observatories are captured at 30 frames per second. Additionally, FRIPON cameras have a target spacing of 80\,km compared to the 100-150\,km of the higher-resolution DSLR systems \citep{howie2017build,colas2020fripon}.
Relying on the law of large numbers, this tighter station spacing and increased sampling rate were found to adequately reduce the uncertainties on the trajectories for FRIPON \citep{jeanne2019calibration}. FRIPON's unique approach, which utilises a dense network of lower-resolution cameras capturing at higher frame rates, enables effective triangulation and photometric measurements across a wide field of view, despite the lower resolution of individual cameras.

\subsection{Data processing}
This study focused on Geminid meteor shower FRIPON observations during 2016-2023 (584 fireballs). The goal is to check the differences in results between the data reduction pipelines of FRIPON, DFN, AMOS, and WMPL (as summarised in Table~\ref{tab:velocity_comparison}). The FRIPON pipeline was initially used to process the observations and make the astrometric reductions. For a description of the astrometric reduction methodology of FRIPON, please refer to \citet{jeanne2019calibration}. Afterwards, the single-station observations were converted to the GFE\footnote{\url{https://github.com/UKFall/standard}} standard format for processing by the three other data processing pipelines \citep{rowe2021just}. Astrometric calibration is highly network–specific and cannot realistically be rerun in four independent codes for this comparison. DFN solves a single long–exposure DSLR frame that contains $\sim10^{3}$ reference stars \citep{howie2017build}; WMPL/RMS derives solutions from 88\degr$\times$48\degr\ HD video with typically 50–200 stars per stack \citep{vida2021global}; whereas low-resolution FRIPON cameras must integrate a dedicated 5-s exposure and merge a few dozen detections from many epochs to reach a $\gtrsim100$-star solution \citep{jeanne2019calibration}. AMOS performs all-sky video astrometry based on the procedure of \citet{1995A&AS..112..173B} with typically 100–300 stars. Porting these four very different astrometry strategies into a single cross-network astrometric pipeline would be a major software project and is beyond the scope of the present work; instead, we rely on the published FRIPON solutions (median $\simeq1$ arc-min residuals; \citealp{jeanne2019calibration}) as a common, quality-controlled baseline for the trajectory and orbit comparison that follows.

Several methods have been developed to identify meteor showers and assess orbital similarity. The earliest was the $D$-criterion introduced by \citet{southworth1963statistics}, which calculates orbital similarity based on the orientation ($i$, $\omega$, $\Omega$), perihelion ($q$), and shape ($e$) of the orbits. Improvements to this approach have been proposed by \citet{drummond1981test}, \citet{jopek1993remarks}, and \citet{jopek2008meteoroid}. \citet{jenniskens2008meteoroid}, on the other hand, proposed a criterion rooted in dynamical arguments, which evaluates the likelihood of a meteoroid stream’s association with a parent body by considering how orbital evolution due to secular perturbations aligns the orbits over time. 

In this study, we applied a distance function, $D_N$, based on four geocentric quantities directly linked to observational data, as proposed by \citet{valsecchi1999meteoroid}. Unlike conventional orbital similarity criteria, which rely on comparing derived orbital elements, the $D_N$ method operates in a multidimensional space defined by the independently measured physical quantities. It incorporates components of the geocentric velocity at the Earth encounter and introduces two near-invariant variables that account for the effects of secular perturbations on meteoroid orbits. This focus on directly observable quantities offers a more robust approach to identifying meteor showers, minimising reliance on derived parameters that can be subject to larger uncertainties.

To ensure reliability, a threshold of $D_N < 0.1$ was chosen, as \citet{shober2024generalizable} demonstrated that this value maintains a false-positive rate $<$ 2.5\% for EN meteor showers. Consequently, assuming similar statistics, more than 97.5\% of the meteors identified here should be confidently classified as Geminids.

\section{Results}\label{sec:results}

\begin{table*}[ht]
    \caption{Summary table for the nominal median measurements of the Geminid meteor shower.}
    \label{tab:geminids}
    \centering
    \small
    \begin{threeparttable}
    \begin{tabularx}{\textwidth}{@{} l 
        S[table-format=3.1] S[table-format=3.1] 
        S[table-format=3.1] S[table-format=2.1] 
        S[table-format=2.1] S[table-format=1.2] 
        S[table-format=1.2] S[table-format=1.3] 
        S[table-format=1.3] S[table-format=3.1] 
        S[table-format=3.1] S[table-format=2.1] 
        X X X @{}}
    \toprule
        \textbf{Activity} & \textbf{RA} & \textbf{$\sigma_{RA}$} & \textbf{Dec} & \textbf{$\sigma_{Dec}$} &\boldmath{$v_{g}$} & \textbf{$\sigma_{v_{g}}$} &\textbf{a} & \textbf{q} & \textbf{e} & \textbf{Peri} & \textbf{Node} & \textbf{Inc} & \textbf{N} & \textbf{Ref} & \textbf{Obs} \\ 
    \midrule
        annual & 112.9 & {\textemdash} & 32.3 & {\textemdash} & 34.6 & {\textemdash} & 1.38 & 0.141 & 0.898 & 324.2 & 261.6 & 23.5 & 51 & A & photo. \\  
        annual & 113.8 & {\textemdash} & 32.3 & {\textemdash} & 34.6 & {\textemdash} & 1.40 & 0.14  & 0.897 & 324.4 & 262.2 & 23.9 & 279 & B & photo. \\  
        2002-2006 & 112.8 & {\textemdash} & 32.1 & {\textemdash} & 35.0 & {\textemdash} & 1.416 & 0.136 & 0.904 & 324.6 & 261.3 & 24.0 & 4384 & C & radar \\  
        2007-2008 & 112.8 & {\textemdash} & 32.3 & {\textemdash} & 33.5 & {\textemdash} & {\textemdash} & {\textemdash} & {\textemdash}& {\textemdash}& {\textemdash} & {\textemdash} & 2510 & D & video \\  
        2002-2008 & 112.5 & {\textemdash} & 32.1 & {\textemdash} & 34.5 & {\textemdash} & 1.35 & 0.1373 & 0.898 & 324.95 & 261.0 & 23.2 & 10381 & E & radar \\  
        2010-2013 & 113.5 & {\textemdash} & 32.3 & {\textemdash} & 33.8 & {\textemdash} & 1.31 & 0.145 & 0.889 & 324.3 & 261.7 & 22.9 & 5103 & F & video \\  
    \midrule
        2016-2023 & 113.6 & 0.73 & 32.3 & 0.44 & 34.0 & 0.41 & 1.32 & 0.144 & 0.891 & 324.3 & 262.3 & 23.2 & 584 & FRIPON & video \\  
        2016-2023 & 113.9 & 0.74 & 32.3 & 0.48 & 33.6 & 0.60 & 1.29 & 0.147 & 0.886 & 324.2 & 262.3 & 22.6 & 584 & DFN & video \\  
        2016-2023 & 113.9 & 0.75 & 32.3 & 0.51 & 33.8 & 0.56 & 1.30 & 0.145 & 0.888 & 324.3 & 262.3 & 22.9 & 584 & AMOS & video \\  
        2016-2023 & 113.9 & 0.75 & 32.3 & 0.42 & 33.8 & 0.47 & 1.30 & 0.145 & 0.888 & 324.3 & 262.3 & 22.8 & 584 & WMPL & video \\  
    \bottomrule
    \end{tabularx}

    \tablefoot{The table compares values from previous studies and the 2016-2023 FRIPON observations processed through four separate data reduction pipelines. Radiant and orbital information is provided using the same nomenclature as the IAU-MDC, and units are given in au, degrees, and km\,s$^{-1}$. The letter next to the reference indicates the radiant in Fig.~\ref{fig:geminid_radiants}. The last four rows are based on the same 584 Geminid fireballs observed by the FRIPON network, where the $\sigma$ values represent the estimated standard deviation for the distribution of Geminids, rather than an uncertainty of an individual measurement.}

    \begin{tablenotes}
      \footnotesize
      \begin{minipage}{\textwidth}
        \begin{multicols}{2}
            \item[A] \citep{jopek2003meteor} (Harvard meteors)
            \item[B] \citep{jopek2003meteor} (MORP, Dutch Meteor Society, Harvard)
            \item[C] \citep{brown2008meteoroid}
            \item[D] \citep{sonotaco2009meteor}
            \item[E] \citep{brown2010meteoroid}
            \item[F] \citep{jenniskens2016established}
        \end{multicols}
      \end{minipage}
    \end{tablenotes}

    \end{threeparttable}
\end{table*}

The Geminid meteor shower, known for its robust and consistent activity in mid-December originating from 3200 Phaethon, has been extensively monitored by the FRIPON network from 2016 to 2023. Here, we have compiled and analysed observations of 584 FRIPON-observed Geminid fireballs. This dataset provides a substantial basis for evaluating the efficacy and consistency of four distinct meteor data reduction pipelines (FRIPON, DFN, AMOS, and WMPL) while also providing a new analysis of the Geminid meteor shower. Furthermore, the results can also be compared against historically established values for the radiant, velocity, and orbital elements of the Geminids estimated by five separate studies completed over the last 20 years and documented by the IAU Meteor Data Center\footnote{\url{https://www.ta3.sk/IAUC22DB/MDC2022/}}.

Table~\ref{tab:geminids} summarises the resulting median values from the four pipelines alongside established values from the aforementioned studies, providing a comprehensive comparison across multiple methodologies. This table displays a general consistency of the radiant and velocity estimates obtained from the different pipelines with prior recorded data. Although the pipeline results are generally congruent with the established records, there are discernible, albeit minor, systematic variations attributed to each pipeline. To visualise all pairwise differences between pipelines, we provide compact $\Delta$–matrix histograms in Appendix~\ref{sec:appendix} for radiants (Figs.~\ref{fig:app_dRA}, \ref{fig:app_dDec}), geocentric speed (Fig.~\ref{fig:app_dvg}), and orbital elements (Figs.~\ref{fig:app_da}–\ref{fig:app_dOmega}).

\subsection{Velocities}
Previous velocity estimates for the Geminids meteor shower, as shown in the top six rows of Table~\ref{tab:geminids}, report a 1.5\,km\,s$^{-1}$ range of nominal median values for $v_{g}$. Typically, the video observations report the slowest speeds (33.5-33.8\,km\,s$^{-1}$; \citealp{sonotaco2009meteor,jenniskens2016established}), linked to previous findings that video observations tend to underestimate Geminids velocities \citep{hajdukova2017meteoroid}. Photographic observations of the Geminids tend to be slightly faster (34.6\,km\,s$^{-1}$; \citealp{jopek2003meteor}); however, the IAU Meteor Data Center (MDC) only lists one study that uses 51 older, legacy Harvard photographic plate observations (1940s-1950s), along with five Meteorite Observation and Recovery Project (MORP) observations (1971-1985), and 223 photographic observations by the Dutch Meteor Society (1972-96). More recent photographic EN observations of Geminid fireballs report slightly lower speeds ($\sim$34.1\,km\,s$^{-1}$;\citealp{borovivcka2022_one,henych2024mechanical}). Finally, radar observations of the Geminids have the highest velocities (34.5-35.0\,km\,s$^{-1}$; \citealp{jopek2003meteor,brown2010meteoroid}).

\begin{figure}
    \centering
    \includegraphics[width=\linewidth]{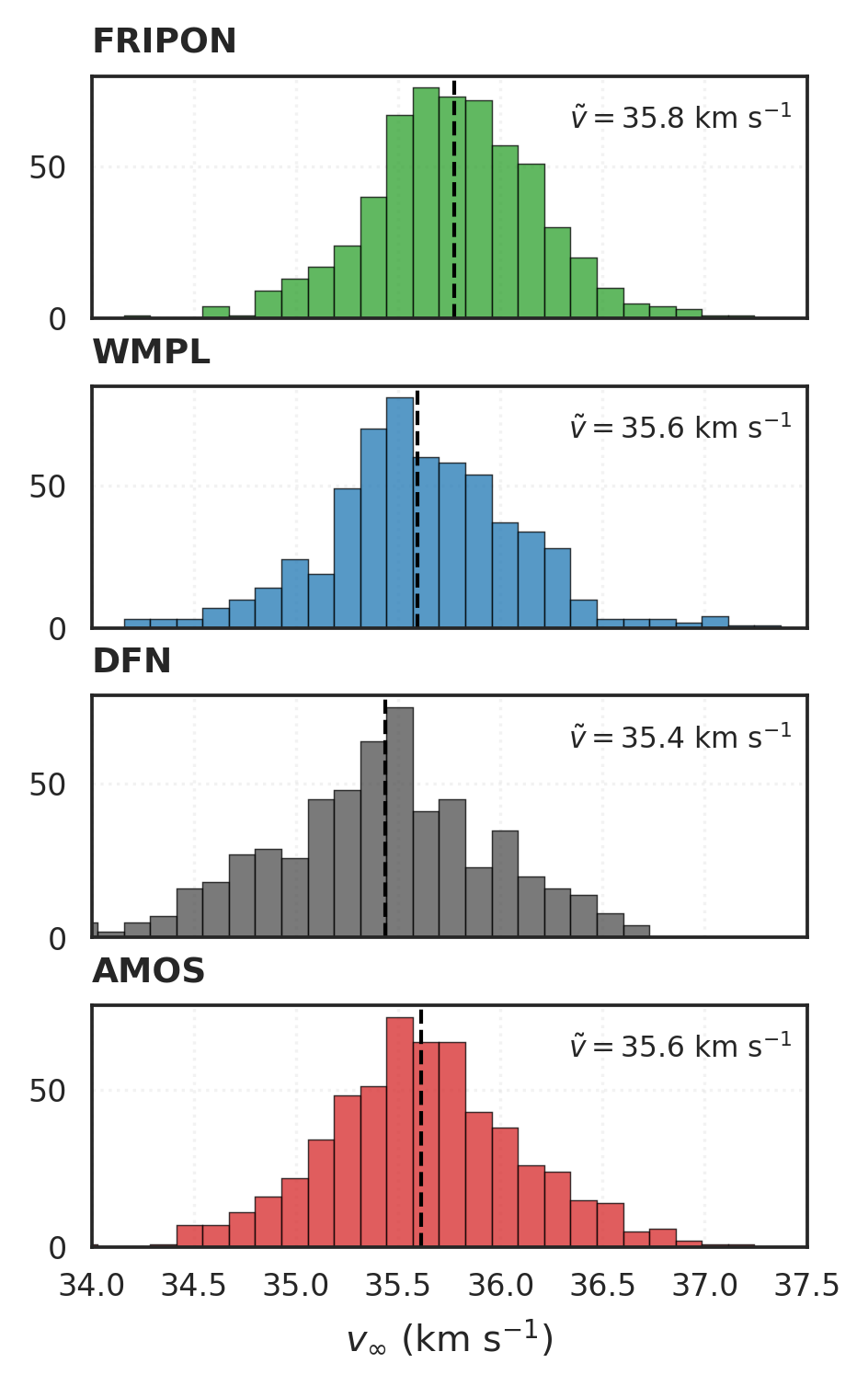}
    \caption{Distribution of $v_{\infty}$ corrected for pre-luminous deceleration using the model of \citet{vida2018modelling}. The dashed vertical line and corresponding value correspond to the median $v_{\infty}$.}
    \label{fig:v_inf_hist}
\end{figure}

The four tested pipelines provide an estimate of the velocity in the lower range of previously published values.
The maximum velocity difference is $1.4$\,km\,s$^{-1}$($35.0$\,km\,s$^{-1}$ from \citet{brown2008meteoroid} and $33.6$\,km\,s$^{-1}$ from DFN - this work). However, the lowest nominal velocity of the shower is by \cite{sonotaco2009meteor}, with a given velocity of $33.5$\,km\,s$^{-1}$.
The AMOS and WMPL data reduction pipelines yield very similar median geocentric velocities ($\sim$33.8\,km\,s$^{-1}$), while the DFN's median value is slightly less (33.6\,km\,s$^{-1}$) and FRIPON is more (34.0\,km\,s$^{-1}$). While relatively minor, such differences could be crucial for accurate stream modelling and impact the interpretation of expected activity. As seen in Fig.~\ref{fig:v_inf_hist}, where pre-luminous deceleration corrections are applied to all $v_{\infty}$ estimates, the deviations between the estimated $v_{\infty}$ values of the four pipelines persist. In other words, the velocity deviations are not explained by pre-luminous deceleration and likely stem from the triangulation or velocity solver used (Table~\ref{tab:velocity_comparison}). Assuming an asteroidal composition observed by a FRIPON-like system, the median estimated pre-luminous deceleration is only 25\,m\,s$^{-1}$ (Fig.~\ref{fig:pre_luminous_hist}). The corresponding pairwise distributions of $\Delta v_{g}$ are provided in Appendix~\ref{sec:appendix} (Fig.~\ref{fig:app_dvg}).

\begin{figure}
    \centering
    \includegraphics[width=\linewidth]{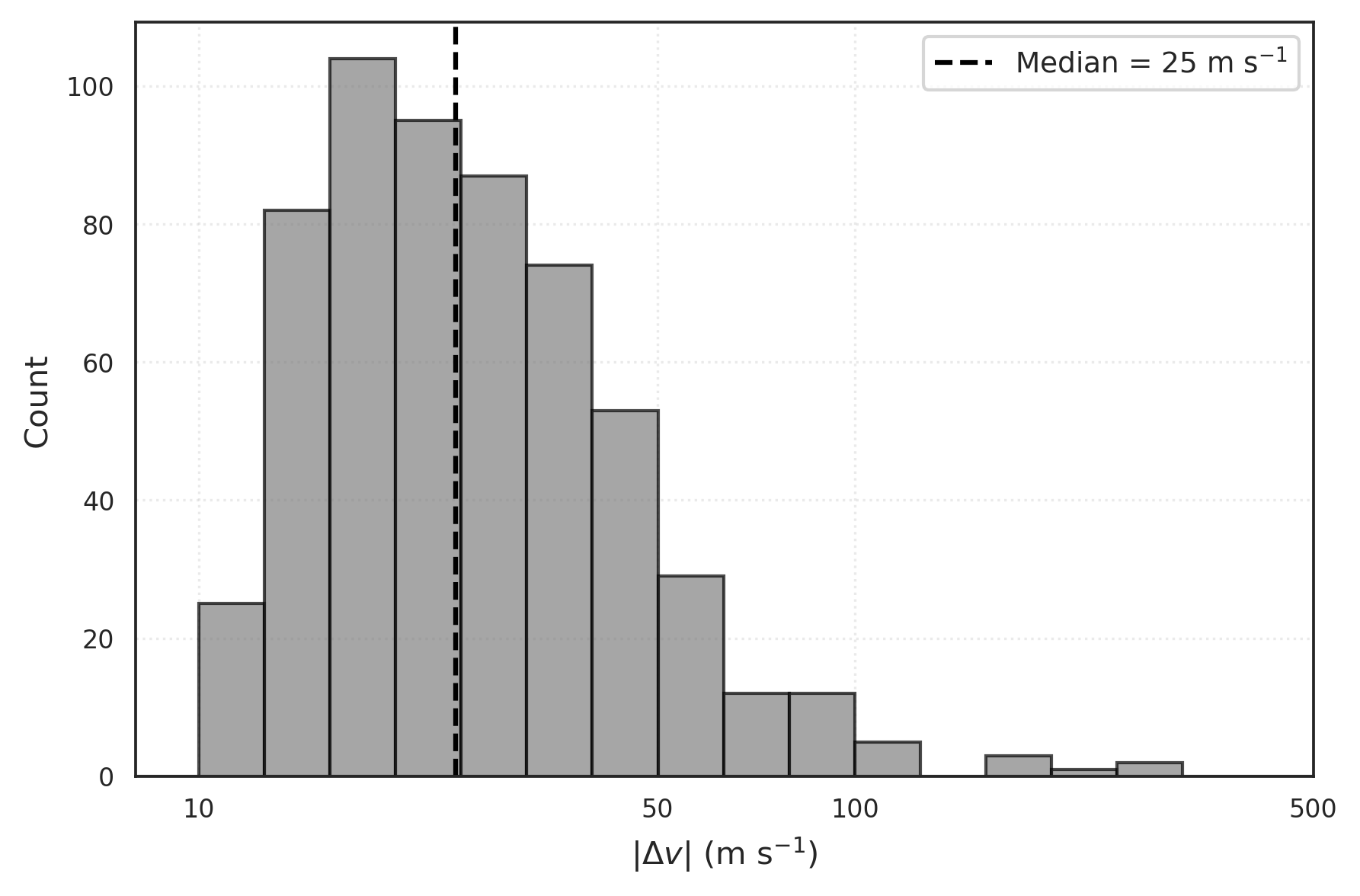}
    \caption{Pre-luminous deceleration estimated for the 584 Geminid FRIPON fireballs using the model of \citet{vida2018modelling}, showing that $>$98\% experienced less than 100\,m/s of deceleration before detection. The dashed vertical line indicates the median value.}
    \label{fig:pre_luminous_hist}
\end{figure}

\begin{figure*}[]
\centering
\includegraphics[width=\linewidth]{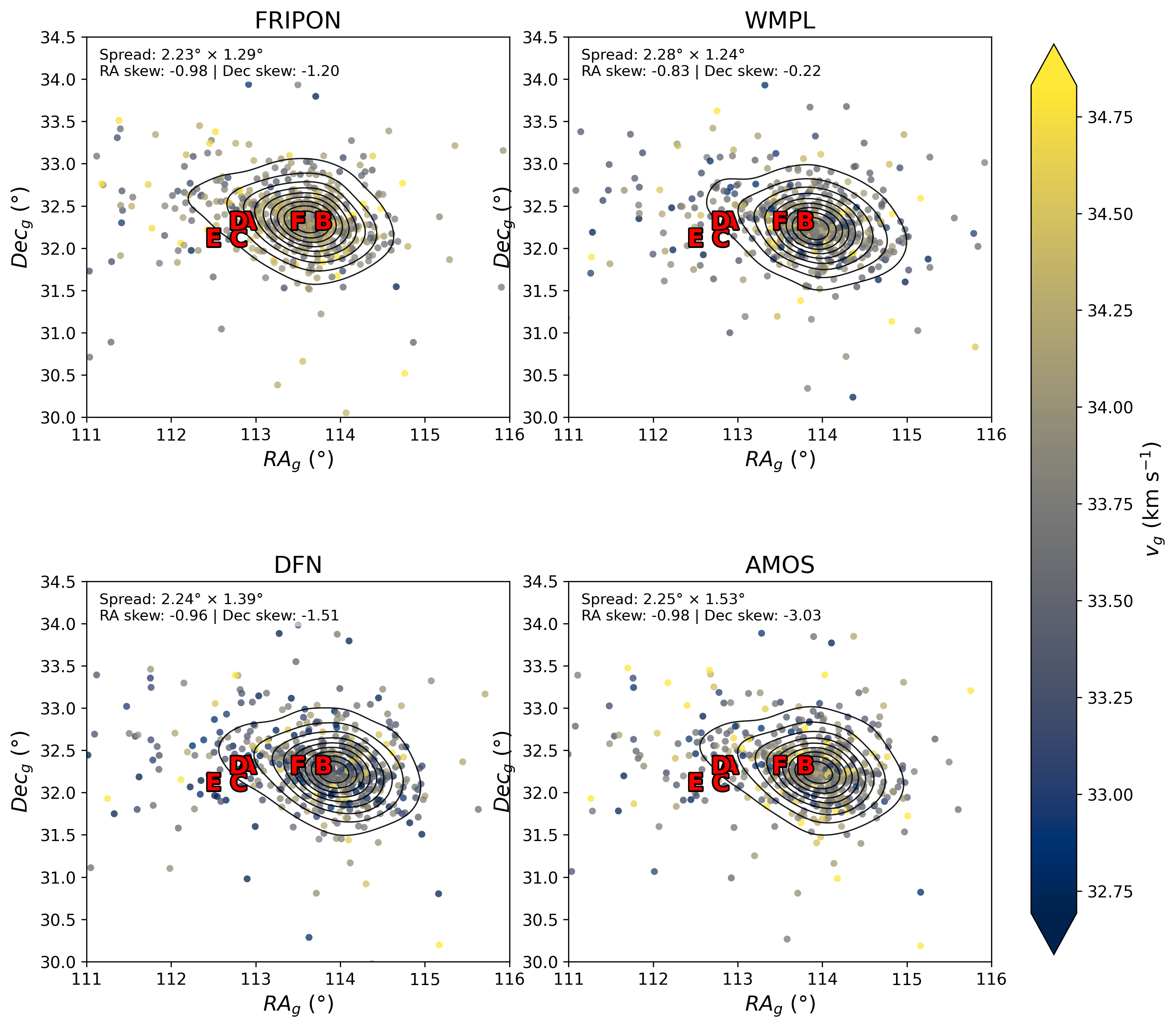}
\caption{Calculated geocentric radiant distribution for the Geminid meteor showers based on FRIPON observations between 2016-2023. The six black points with crosses indicate radiants identified from previous studies of the Geminids (\textbf{A:} \citealp{jopek2003meteor}, \textbf{B:} \citealp{brown2008meteoroid}, \textbf{C:} \citealp{sonotaco2009meteor}, \textbf{D:} \citealp{brown2010meteoroid}, \textbf{E:} \citealp{jenniskens2016established}). Nominal values for the four pipelines and previous results can also be found in Table~\ref{tab:geminids}.}
\label{fig:geminid_radiants}
\end{figure*}

\begin{figure}
    \centering
    \includegraphics[width=\linewidth]{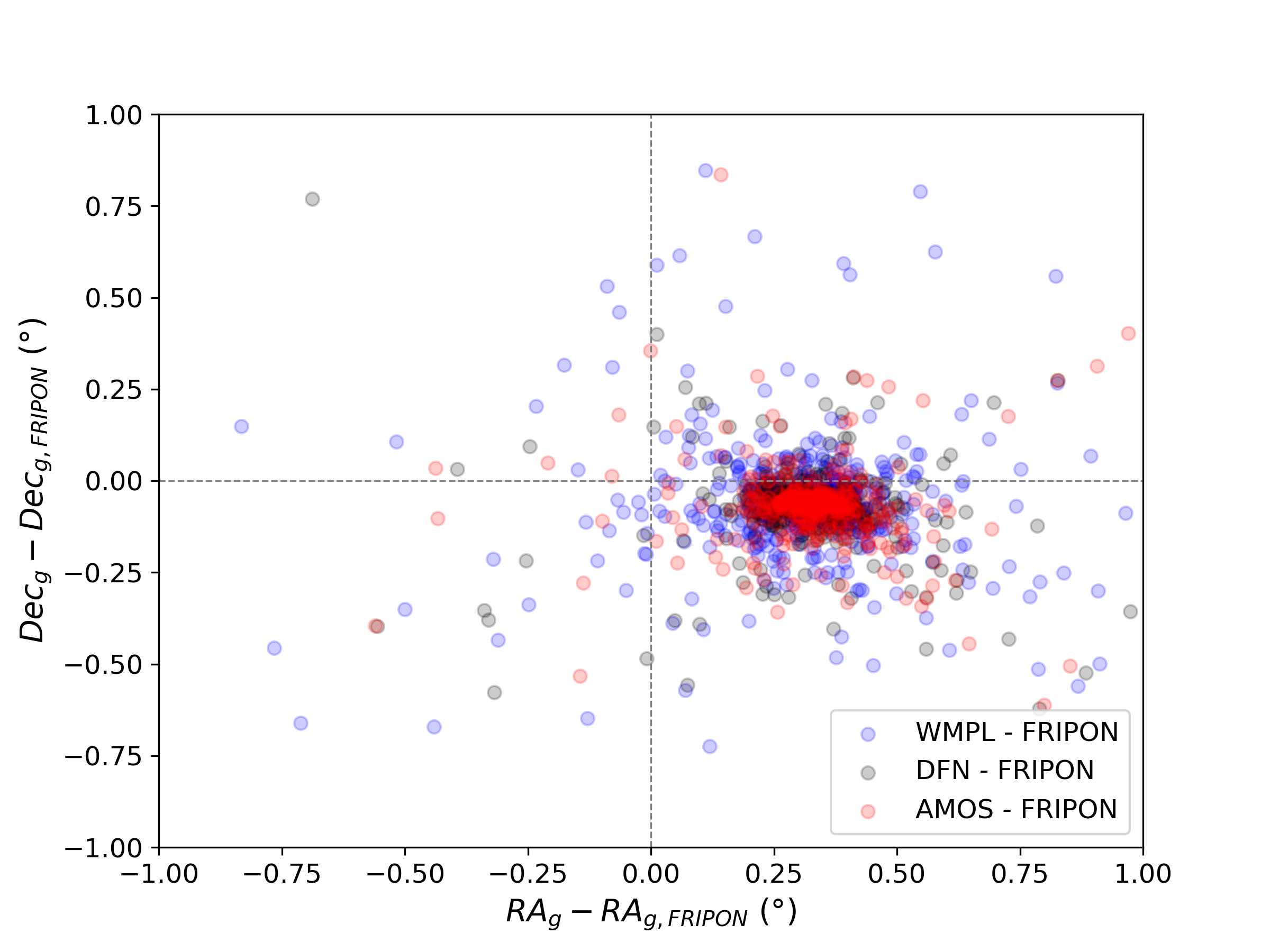}
    \caption{Variation in the FRIPON-observed Geminid fireball radiants calculated by the DFN, AMOS, and WMPL data reduction pipelines relative to those calculated by the FRIPON automated pipeline.}
    \label{fig:geminid_radiant_variation}
\end{figure}

\subsection{Radiants}
Fig.~\ref{fig:geminid_radiants} showcases the calculated radiant distributions, which align closely with those recorded in earlier studies, such as those by \citet{jopek2003meteor} (MORP, Dutch Meteor Society, Harvard) and \citet{jenniskens2016established}. Very slight deviations are observed, particularly in the right ascension values. The nominal right ascension values reported by the four other previous studies (A, C, D, E in Table~\ref{tab:geminids}) are notably smaller than any of the distributions produced here by the four different pipelines, providing strong evidence that this difference is real for the FRIPON Geminid dataset. As illustrated in Fig.~\ref{fig:geminid_radiant_variation}, there is also a discernible variation in the computed radiants when comparing the results from different pipelines relative to those calculated by the FRIPON automated system. The DFN, AMOS, and WMPL radiants are all almost identical, displaying only slightly lower declinations (-0.1$^{\circ}$) but much larger right ascensions (+0.3$^{\circ}$) relative to the radiants calculated by FRIPON's pipeline. Full pairwise comparisons for $\Delta$RA and $\Delta$Dec across all pipelines are shown in Appendix~\ref{sec:appendix} (Figs.~\ref{fig:app_dRA} and \ref{fig:app_dDec}). 

To comprehensively analyse the distribution of radiants, we also quantified both the spread and skew of the geocentric radiant measurements. The spread was characterised using Principal Component Analysis (PCA) of the two-dimensional (RA, Dec) distribution, allowing us to identify the semi-major and semi-minor axes of the distribution ellipse. These axes correspond to the square root of the eigenvalues of the covariance matrix, scaled to represent a 3\,$\sigma$ dispersion along the directions of greatest and least variance. Skewness was computed separately for RA and Dec to quantify any asymmetry in the distribution around its mean. Positive skew indicates a longer tail to the right (higher values), while negative skew indicates a longer tail to the left (lower values). Symmetric distributions exhibit skewness near zero. 

All four pipelines produce similar radiant distributions and ranges, with standard deviations for RA all concentrating around \(0.74^\circ\) and Dec ranging from \(0.43^\circ\) to \(0.51^\circ\). The 3\,$\sigma$ PCA spread is found to be generally around (\(2.2^\circ \times 1.3^\circ\)). Regarding skewness, all pipelines show similar moderate negative skew in RA and Dec, indicating a slight asymmetry with a bias toward lower RA and Dec values. The RA skew values range between \(-1.0\) to \(-0.8\), and Dec skew with a larger diversity from \(-3.03\) to \(-0.20\). Dec skewness is more pronounced for AMOS (\(-3.0\)), primarily due to a small number of outliers. 

Additionally, in accordance with several previous works, we have also calculated the median absolute angular offset from the mean radiant for each pipeline. These offsets are \(0.45^\circ\), \(0.47^\circ\), \(0.45^\circ\), and \(0.48^\circ\) for FRIPON, DFN, WMPL, and AMOS, respectively. For reference, \citet{moorhead2021meteor} found a median offset of \(0.38^\circ\)) for 1279 Geminids observed by GMN and \citet{kresak1970dispersion} found a \(0.49^\circ\)) value based on 82 observations. 

\begin{table}[t]
\centering
\caption{One–sigma single radiant measurement uncertainties for the FRIPON Geminids sample.}
\begin{tabular}{lccccc}
\hline
\multirow{2}{*}{Pipeline} &
\multicolumn{2}{c}{$\sigma_{\mathrm{RA}}$ (deg)} &
\multicolumn{2}{c}{$\sigma_{\mathrm{Dec}}$ (deg)} \\
\cline{2-3}\cline{4-5}
 & Median & 95\,\% range & Median & 95\,\% range \\
\hline
FRIPON & 0.117 & 0.030--0.372 & 0.117 & 0.030--0.372 \\
WMPL   & 0.067 & 0.015--0.462 & 0.053 & 0.011--0.461 \\
DFN    & 0.030 & 0.013--0.083 & 0.021 & 0.011--0.064 \\
AMOS   & 0.073 & 0.024--0.763 & 0.062 & 0.023--0.550 \\
\hline
\end{tabular}
\tablefoot{The “95\,\% range’’ column encloses the central 95\,\% of the distribution
(low–high). This range is distinct from the larger range of nominal radiant values (as seen in Fig.~\ref{fig:geminid_radiants}).}
\label{tab:radiant-uncertainties}
\end{table}

\begin{figure}
    \centering
    \includegraphics[width=\linewidth]{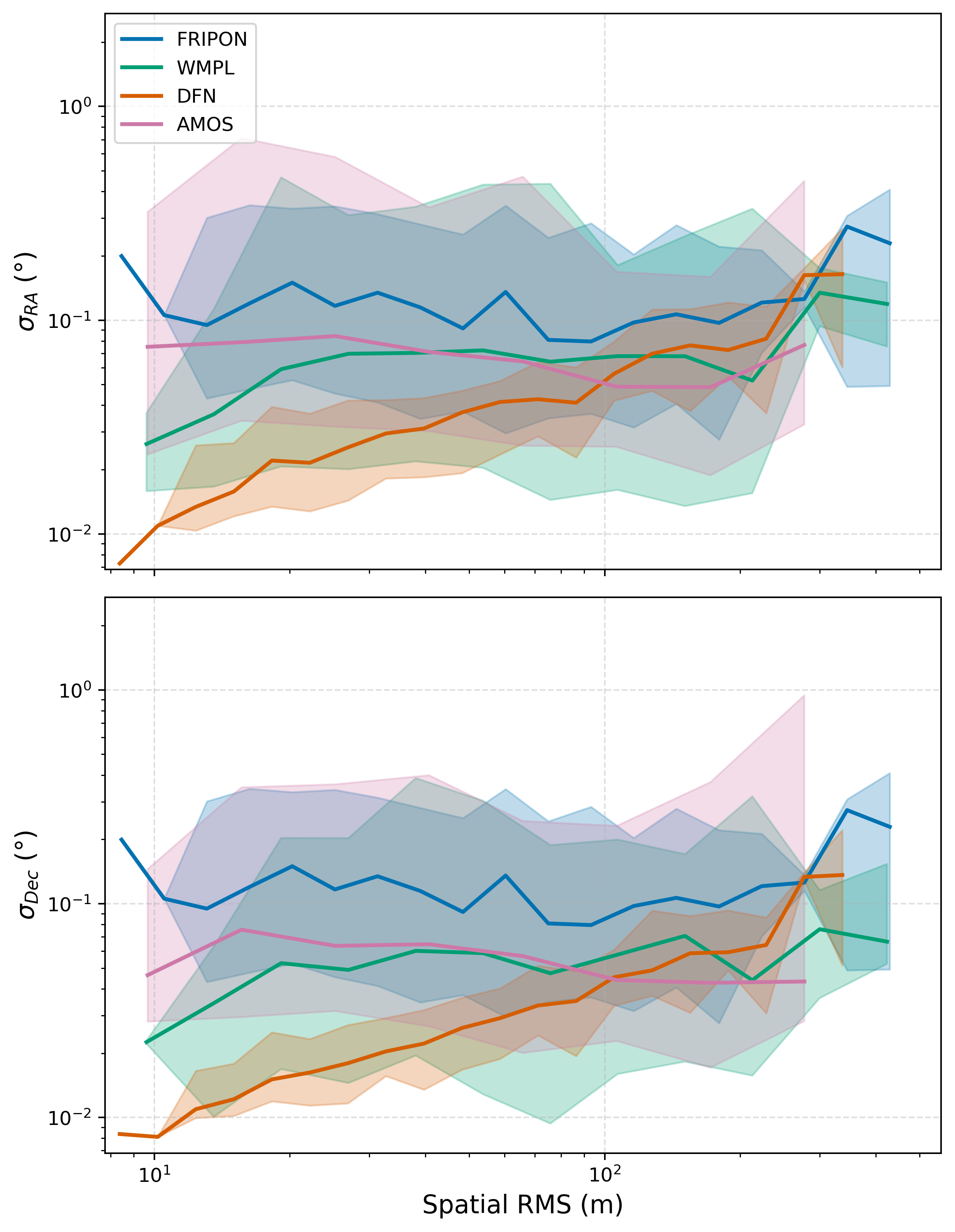}
    \caption{Median \(1\sigma\) radiant uncertainties in (top) right ascension and (bottom) declination as a function of the trajectory-fit cross-track RMS residuals.  
    Values are binned logarithmically in RMS; the solid line in each colour-coded curve is the median of all events in the bin, while the shaded region encloses the 5\,--\,95\,\% percentile range (central 90\,\%) and illustrates the event-to-event spread. FRIPON, WMPL, and AMOS pipelines (nearly flat curves) report uncertainties that are largely independent of fit quality. In contrast, the DFN solution shows a clear positive trend, indicating that its formal errors scale with the residuals.}
    \label{fig:rms_radiant_err}
\end{figure}

All four pipelines deliver similar radiant precision; however, the DFN pipeline's 95\% confidence region is the smallest (Table~\ref{tab:radiant-uncertainties}). The median $\sigma_{RA}$ and $\sigma_{Dec}$ cluster tightly around 0.02–0.11\,$^{\circ}$. Fig.~\ref{fig:rms_radiant_err} compares 1-$\sigma$ radiant uncertainties against the cross-track RMS. DFN is the only pipeline whose uncertainties scale strongly with the RMS (Spearman $\rho \approx $0.7), meanwhile, the other pipelines show only weak correlations ($\rho<$0.2). This occurs because, for this FRIPON comparison only, we supplied the DFN pipeline with a single, fixed measurement noise value for every pick; when the residuals exceed this assumption, the EKF inflates the model covariance, resulting in larger radiant $\sigma$ values. In contrast, WMPL, AMOS, and FRIPON weight astrometric noise on a per-pick basis, so their $\sigma$-to-RMS correlations remain weak ($\rho<$0.2). Nevertheless, the vast majority of individual uncertainties are well below the observed radiant dispersion of the Geminid stream (Fig.~\ref{fig:geminid_radiants}), thus confirming the capability to discern substructure from FRIPON observations.

\subsection{Velocity Uncertainties}
A systematic difference emerges in the reported velocity uncertainties across the networks. This variation in uncertainties is clearly illustrated in Fig.~\ref{fig:geminid_rms_v_err}, where one-sigma velocity uncertainties are plotted against the spatial RMS of the cross-track residuals of the trajectories fitted by each pipeline. Although all pipelines exhibit comparable RMS values, indicating similar goodness-of-fit in trajectory modelling, the derived velocity uncertainties differ markedly. In particular, the WMPL pipeline consistently reports the lowest uncertainties, followed by AMOS, then FRIPON, with the DFN pipeline exhibiting the highest uncertainties despite similar, nearly identical RMS distributions. 

These findings underscore a critical challenge in meteor trajectory analysis: while multiple pipelines can achieve similar trajectory fits, their approaches to estimating velocity uncertainties diverge significantly. Since accurate uncertainties are essential for reliable orbital characterisation, this inconsistency affects subsequent interpretations of meteoroid stream dynamics. Therefore, discussing uncertainty quantification across pipelines is crucial for improving the robustness and comparability of meteor observations and derived orbital parameters.

\section{Discussion}\label{sec:discuss}
Our comparative analysis across the four pipelines -- FRIPON, DFN, AMOS, and WMPL -- indicates that while the nominal radiant and velocity estimates derived from DFN, AMOS, and WMPL are largely consistent, the FRIPON pipeline shows subtle but notable systematic deviations. Specifically, the FRIPON radiant exhibits a right ascension approximately 0.3$^\circ$ lower than the median 113.9$^\circ$ reported by DFN, AMOS, and WMPL, accompanied by a marginally larger declination distribution. Additionally, FRIPON tends to produce geocentric velocities that are 300–500\,m\,s$^{-1}$ higher than the medians provided by the other pipelines. Although the spread and skew of the velocity distributions among DFN, AMOS, and WMPL are similar, the velocity uncertainty estimates vary systematically across all networks, with DFN reporting the largest uncertainties and WMPL the smallest. These small systematic offsets propagate into the orbital elements; the pairwise $\Delta a$, $\Delta e$, $\Delta i$, $\Delta\omega$, and $\Delta\Omega$ distributions are summarised in Appendix~A (Figs.~\ref{fig:app_da}–\ref{fig:app_dOmega}).

\subsection{Outliers}
As shown in Table 1, the FRIPON network employs a fully automated data reduction pipeline designed to minimise meteorite recovery times. Candidate events are detected, matched across stations, and the trajectory is solved without human intervention \citep{jeanne2019calibration,colas2020fripon,2023MNRAS.518.2810A}. The survey is optimised for bright, multi-station fireballs (peak magnitude $M_{\!p}<0$), thus, each event is typically recorded by several cameras. This redundancy, together with built-in pixel-level vetoes for slow artefacts (e.g., the Moon, street lamps, aircraft, etc.), means that most spurious picks are either never linked into an event or are down-weighted during the least-squares fit of Eq.~\eqref{eq:friponchi2}. In practice, only a few percent of frames have to be rejected, and the residual false detections have a negligible influence on the final bright‐flight line.

In contrast, the DFN does not automate the fireball picking step, leading to precise trajectories and no outliers. Further manual refinement is also done for potential meteorite-dropping events. Thus, the DFN pipeline does not include outlier detection by design. As a result, when initially fed FRIPON data containing false detections and large residuals, the DFN solver often fails, since it is designed to raise errors when it encounters large residuals during trajectory determination. Similarly, the AMOS pipeline encountered difficulties with FRIPON data. Many FRIPON observations required manual cleaning before AMOS could process them. When bad data were not removed, the resulting trajectories and orbital solutions deviated significantly, underscoring the importance of effective outlier rejection. These issues highlight the need for rigorous preprocessing of fireball observation files as well as the tighter radiant and velocity distribution output by the FRIPON pipeline.

\begin{figure}
    \centering
    \includegraphics[width=\linewidth]{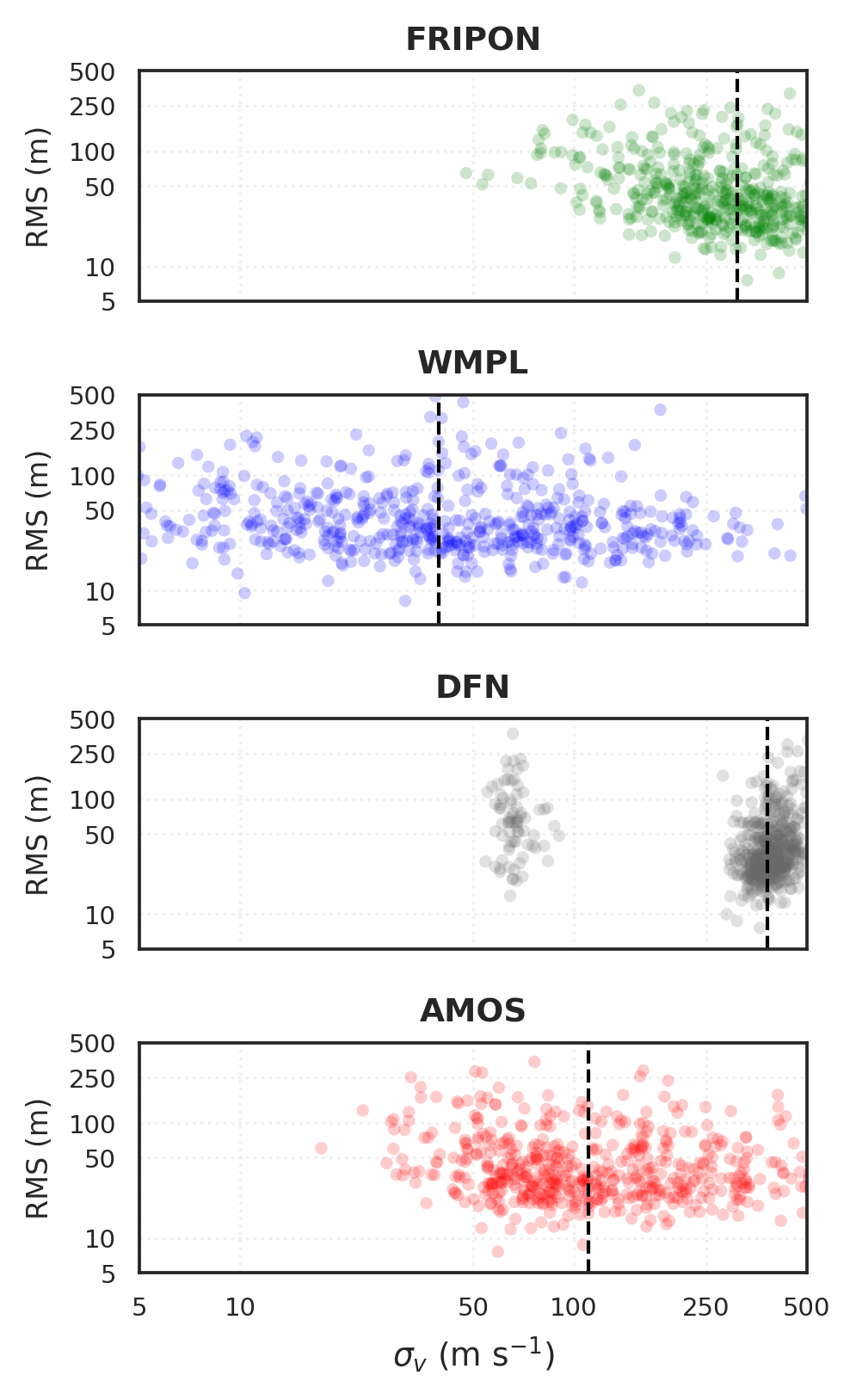}
    \caption{Variation in the RMS of the cross-track spatial residuals (m) for the trajectory fitting versus the velocity uncertainty estimated by each pipeline. Vertical dashed lines indicated the median value. All pipelines have relatively equal RMS ranges; however, the velocity uncertainties differ systematically.}
    \label{fig:geminid_rms_v_err}
\end{figure}

\begin{figure}
    \centering
    \includegraphics[width=\linewidth]{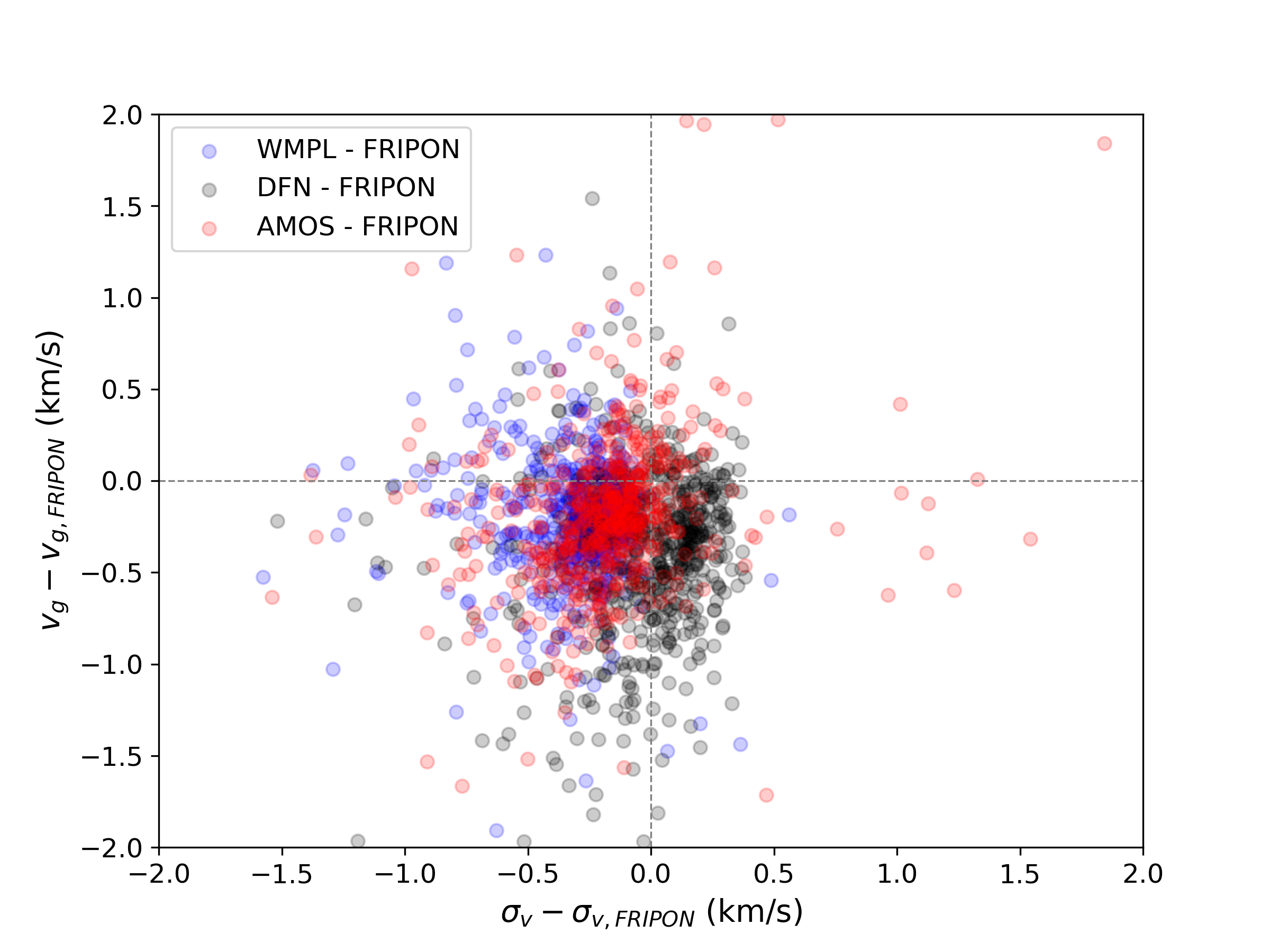}
    \caption{Comparison of the velocity and velocity uncertainty of Geminid meteor shower FRIPON events processed by the DFN, AMOS, WMPL pipelines versus the automatic reduction pipeline of FRIPON. The x-axis is the difference between the $\sigma_{v_{g}}$ values processed by the DFN (a), AMOS (b), or WMPL (c) and those calculated by FRIPON's pipeline. The y-axis similarly shows the differences for the nominal $v_{g}$ values.}
    \label{fig:geminid_v_variation}
\end{figure}

\subsection{Radiant and Reference Frames}
FRIPON’s astrometric pipeline is anchored to an inertial celestial frame. It uses the ICRF2/J2000 celestial reference frame for star calibration \citep{colas2020fripon}. Stars from the Hipparcos catalogue (an ICRF/J2000 reference) are matched in each camera’s field to determine orientation. In practice, FRIPON fits the camera model by converting star coordinates from the celestial frame to the Earth-fixed frame (ITRF) at the time of observation \citep{jeanne2019calibration}. This implies that FRIPON does account for Earth’s rotation and precession when mapping star positions – essentially using star coordinates precessed to the observation epoch during calibration. The intent is to produce meteor directions in a J2000-equatorial frame for consistency \citep{colas2020fripon}. However, if the final radiant coordinates are not explicitly transformed back to the J2000 epoch, a slight epoch mismatch could remain. A $\sim$\,20-year difference between the observation date and J2000 can shift right ascension on the order of 0.3° due to precession. In other words, if FRIPON assumes its results are ``in J2000'' without actually rotating from the date-of-observation equinox, the RA would appear about 0.2$^\circ$–0.3$^\circ$ off (drifting 50.3 arcseconds per year), exactly as observed. Thus, FRIPON’s radiant output is effectively in the epoch of the event (or a recent epoch), whereas the other networks report J2000 coordinates – leading to a systematic RA offset if not corrected for precession.

\subsection{Velocity and uncertainties}
The FRIPON network's tendency to report higher meteoroid velocities likely results from its trajectory-fitting approach and deceleration model (Fig.~\ref{fig:geminid_v_variation}). FRIPON employs a dynamic model of deceleration and mass loss, as described in \citet{jeanne2019calibration} (Eq.~\ref{eq:AB}), which describes the deceleration and ablation of a meteoroid in the atmosphere. 
In practice, the velocity solver overestimates \(V_{\rm e}\) when the observed track shows insufficient resolved deceleration to constrain the deceleration model. 

\begin{figure}
  \centering
  \includegraphics[width=\linewidth]{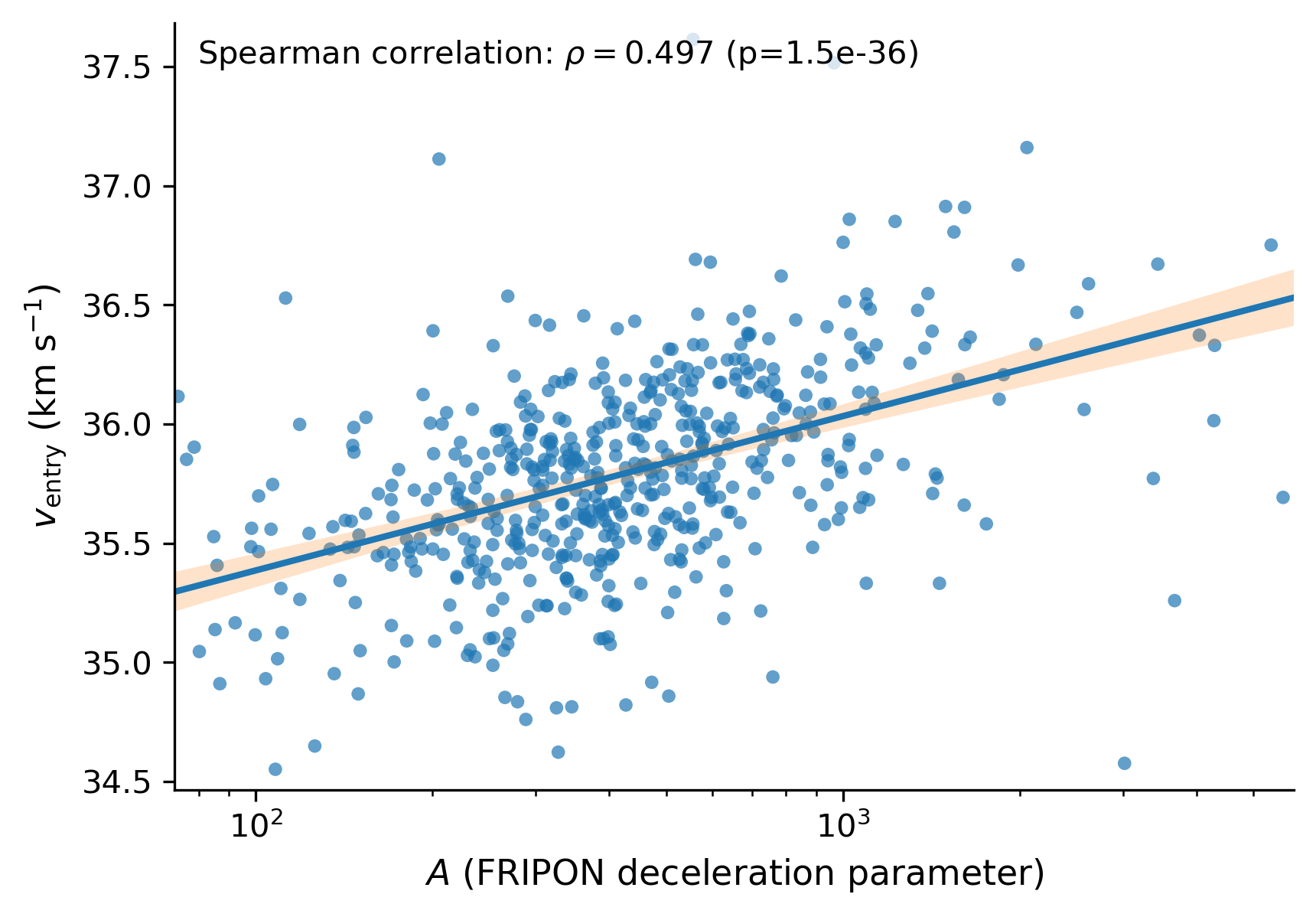}
     \caption{FRIPON-derived entry speed $v_{\mathrm{entry}}$ versus the FRIPON deceleration parameter $A$ \citep{jeanne2019calibration,colas2020fripon}. The solid line is an ordinary-least-squares fit; the shaded band is the 95\% confidence interval. A clear positive trend is evident: meteoroids with larger $A$, i.e., events that display little measurable deceleration, yield systematically higher fitted velocities, indicating that the default FRIPON single-body solver tends to over-estimate the true pre-atmospheric speed when deceleration is weak.}
  \label{fig:alpha_vs_ventry}
\end{figure}

\citet{jeanne2020méthode} (c.f. pages 111-119 within) showed that in the weak-deceleration regime, the separation of \(V_{\rm e}\) from \(A\) becomes ill-conditioned, biasing slightly high \(V_{\rm e}\) and inflating its formal uncertainty. In other words, many $(V_{\mathrm{e}},A)$ pairs fit the luminous segment equally well. Also, if an object ablates away before any deceleration occurs, this can also result in an artificially high A value. So in practice, for short, fast meteors that show minimal deceleration, there's a fundamental ambiguity in the fit that can overestimate velocities and can be identified as a positive correlation between parameters $A$ and $V_{\mathrm{e}}$ (See \citet{jeanne2020méthode} pg. 117).
Fig.~\ref{fig:alpha_vs_ventry} shows a clear correlation between these parameters by the FRIPON pipeline (Spearman $\rho=0.497$, $p=1.5\times10^{-36}$). 

Consequently, velocities for meteorite-dropping fireballs (which decelerate strongly) are expected to be more accurate, whereas the method is not optimal for low-deceleration impacts such as the Geminids examined here.

On the other hand, the lower velocities reported by the DFN pipeline (Fig.~\ref{fig:geminid_v_variation}) are linked to lower-quality observations. As the number of observational data points goes below 50 and the number of cameras is limited to 2, the DFN pipeline tends to produce slightly lower velocities relative to WMPL and AMOS. This is partially true for all the pipelines; however, the DFN was designed for high-precision photographic observations, making it the least equipped to handle noisier observations. This is exemplified by the fact that most of the Geminids analyzed here (321 events), were only observed by two FRIPON cameras. If we only considered events observed by 3+ cameras, the $v_{g}$ estimations start to converge (DFN: 33.7\,km\,s$^{-1}$, WMPL: 33.7\,km\,s$^{-1}$, FRIPON: 34.0\,km\,s$^{-1}$, AMOS: 33.8\,km\,s$^{-1}$). The number of observations significantly influences the data quality of FRIPON events, as the network was not designed with 2-station detections in mind, but rather meteorite-dropping events detected by multiple stations.

Pre-luminous atmospheric deceleration before the first observation can also significantly influence the velocity measurements of meteoroids before their visible ablation phase. \citet{vida2018modelling} demonstrated that the velocity measured at the start of the luminous trajectory is systematically lower than the true pre-atmospheric velocity. This underestimation is particularly significant for smaller or slower meteoroids, where deceleration due to drag and ablation can result in discrepancies ranging from 100 to 750\,m\,s$^{-1}$, depending on the meteoroid’s mass, composition, and initial velocity \citep{vida2018modelling}.  However, this deceleration is not influential within this particular observational dataset, as over 98\% of the meteoroids experienced less than 100\,m\,s$^{-1}$ of pre-luminous deceleration (Fig.~\ref{fig:pre_luminous_hist}), with a median value of 25\,m\,s$^{-1}$. When this extra initial deceleration is taken into account by the DFN, AMOS, and WMPL (Fig.~\ref{fig:v_inf_hist}), the $v_{\infty}$ values barely change. 

Even after pre-luminous deceleration is accounted for, all four data reduction pipelines converge to a lower $v_{g}$ value compared to nearly all other reported values by the IAU MDC (top of Table~\ref{tab:geminids}). Systematic underestimation of Geminid speeds has been reported for several earlier video surveys \citep{hajdukova2017meteoroid}, mainly because those studies (i) centroided faint ($m_{\mathrm{peak}}\gtrsim+2$) meteors and (ii) used straight–line fits that did not correct for deceleration. The four pipelines used here solve for the full dynamical state, and we have accounted for pre-luminous corrections based on the model of  \citet{vida2018modelling}. The higher photographic speed value tabulated in the IAU-MDC derives almost exclusively from a smaller photographic set of \citet{jopek2003meteor}, based on 1940–1950s Harvard plates, a handful of MORP observations (1971-1985), and 223 DMS observations (1972-96); producing a $v_{g}$ estimate that is $\sim800$\,m\,s$^{-1}$ faster than our results. However, more recent modern photographic fireball observations for equally large Geminids recorded by the European Fireball Network produce a median $v_{g}=$\,34.1\,km\,s$^{-1}$ \citep{borovivcka2022_one,henych2024mechanical}. A Kolmogorov–Smirnov test also confirms the difference with EN is significant (\(D=0.38,\;p=4.5\times10^{-5}\)). Thus, there is still a 250-300\,m\,s$^{-1}$ underestimate likely due to instrumental effects. Principally, as pointed out in \citet{hajdukova2017meteoroid}, the automation of point-picking smeared-out motion-blurred meteor streaks can systematically influence the velocity estimates. In particular, this causes the speed to be underestimated when taking the centre of the streak, or overestimated if picking the leading edge \citep{hajdukova2017meteoroid}. The 30-fps frame integration of FRIPON (33 ms) smears a Geminid by \(\sim\)1 km, and centroiding that streak shifts each pick slightly backwards along the path when picking the middle. This should account for the observed 0.25 km s\(^{-1}\) shift relative to EN. In contrast, the much larger MDC discrepancy most likely reflects legacy calibration issues in the Harvard plates rather than a genuine physical difference. 

More notably, as illustrated in Fig.~\ref{fig:geminid_rms_v_err}, although all four data reduction pipelines provide generally similar spatial residual root-mean-square (RMS) values for trajectory fits, there is a clear systematic trend in their estimated velocity uncertainties. The variation in velocity uncertainty outcomes is striking, with the WMPL pipeline yielding the lowest uncertainties, followed by AMOS, FRIPON, and DFN, which reports the highest uncertainty estimates. This discrepancy is not due to poor model fits, as the RMS values for spatial residuals are comparable across all four pipelines at a population scale. Instead, the differences arise from varying methodologies in how each pipeline propagates and estimates uncertainties. 

A factor contributing to differences in reported velocity uncertainties is the treatment of astrometric precision in each pipeline. 
The DFN pipeline is the only method that takes astrometric uncertainties into account and propagates them through to velocity models. These uncertainties are calculated for DFN images within DFN image reduction. FRIPON does not output astrometric uncertainties, though \citet{jeanne2019calibration} notes that FRIPON’s lens calibration routinely achieves $\sim$\,2 arcmin (1$\sigma$) residuals on astrometric measurements.
As the DFN pipeline explicitly requires them (it requires columns $err\_minus\_altitude$, $err\_plus\_altitude$, , $err\_minus\_azimuth$, $err\_plus\_azimuth$ defined as optional in the astrometric standard of https://github.com/UKFAll/standard), to run FRIPON data through the DFN pipeline, we had to set default uncertainties for all observation points. We manually added values of 2 arcmin in azimuth and 4 arcmin in elevation to all observations. 
Including astrometric uncertainties has contributed to the systematically higher velocity uncertainties for the DFN results (Figure \ref{fig:geminid_rms_v_err}). 
However, a more likely factor is the way the EKS method is set up to handle the uncertainties while using a single-body model. As fragmentation is not explicitly modelled by the DFN pipeline, any abrupt deceleration manifests as extra process noise and is absorbed into the covariance matrix. Afterwards, the filter steps down the main-body mass and velocity \citep{sansom2015novel}, resulting in increased overall uncertainties. It is designed to `trust' the observations over the model. 
When combined with the noisier data of FRIPON, an artificially inflated covariance matrix is triggered, explaining the apparent bimodal distribution in Figure (\ref{fig:geminid_rms_v_err}). The lower $\sigma_v$ cloud represents those in `single body' mode, while the higher ones are using the `fragmenting' covariance matrix.
The same two clusters appear within DFN data, confirming that the uncertainty distribution is an intrinsic consequence of the EKS. 
Given these two factors, the larger uncertainties presented here and the large cluster around 300-500\,m\,s$^{-1}$ are a result of the low spatial resolution of the data and the high sampling rate of FRIPON observations, which the EKS is not optimised to handle. While similarly high velocity uncertainties are estimated by the DFN pipeline for DFN observations, they typically constitute a minority subset.

The accuracy of uncertainties has also been studied in \citet{vida2020estimating}, where it was found that a factor of 2-4 underestimation occurs in the radiant uncertainty when 1-$\sigma$ errors are used during the Monte Carlo procedure. This suggests that higher measurement errors should be introduced in the Monte Carlo procedure to ensure accurate uncertainty estimation. However, this analysis assumed that all sources of errors are random and not systematic. \citet{vida2020estimating} demonstrated that systematic biases are also present due to the limitations of measurements (e.g., estimating the initial speed on a segment of the trajectory after some deceleration has already occurred); thus, the problem cannot be collapsed to a one-dimensional consideration of simply inflating the uncertainties. The same effect was also recently noticed by \citet{barghini2025kresaks}, who defined an estimator \(R\) (with \(R\simeq1\) for realistic uncertainties and \(R>1\) when uncertainties are underestimated) and measured \(R=1.19\pm0.05\) for FRIPON but \(R=3.10\pm0.02\) for GMN and CAMS, indicating that the latter two databases publish velocity/radiant errors that are too small near the parabolic limit. This follows the long‑noted link between surplus hyperbolic orbits and underestimated errors \citep[e.g.,][]{hajdukova2024no}, and implies that, while not the smallest, FRIPON’s uncertainties are the most representative of the actual scatter in these observations. Between the four pipelines studied here, the results of \citet{barghini2024recovery} do not necessarily imply that the FRIPON pipeline's uncertainties are the most representative; however, they are less underestimated than the WMPL estimates, according to this recent work.

This highlights a broader issue in the field: a clear disconnect between networks regarding how velocity uncertainties should be assessed. Achieving consensus on uncertainty propagation and how various factors should be accounted for, such as systematics, timing, astrometric precision, and deceleration modelling, is essential for the standardised and reliable understanding of the meteoroid populations in near-Earth space. 

\subsection{Advice for future meteor data creation and recommendations}

Based on our work, we have provided a handful of recommendations: 

\begin{enumerate}
    \item  Be able to output/input observations in the GFE\footnote{\url{https://github.com/UKFall/standard}} format so that data can easily be shared, tested, and combined. 
    \item Include per‑event quality metadata in GFE (e.g., station information, picks, convergence angle, min range, cross‑track RMS, long-track RMS, frame rate, exposure, lens model, clock source, etc.). 
    \item Share the single-station GFE files via Zenodo or some other file-sharing service. 
    \item All data processing pipelines should be publicly available on GitHub, Gitlab, etc., with clear documentation. 
    \item Each network should clearly document the assumed celestial reference frame (e.g., “J2000 equatorial coordinates” or “no-atmosphere topocentric horizontal coordinates”), detailing the reference epoch and refraction correction; all calculations (or final outputs) should be transformed to that frame within arcsecond precision. 
    \item Confirm that each observatory’s time stamps, particularly shutter-opening times or frame mid-exposure times, are accurately and consistently converted to UTC, with verified leap seconds and known latencies/delays. 
    \item Propagate per‑pick astrometry and timing errors through the trajectory and velocity fits, and report per‑event 1$\sigma$ uncertainties and the propagation method (especially for automated point‑picking). 
    \item Specify the spectral sensitivity of the instrument and document the bandpass of the star catalog used for photometric calibration and measurements.
    \item Document any idiosyncrasies of the instrument that will help understand and reproduce the results.
    \item If a group is starting a meteor or fireball network, it is highly recommended to use WMPL\footnote{\url{https://github.com/wmpg/WesternMeteorPyLib}} to process your data. It is well written, publicly available, and well-documented. 
\end{enumerate}

\section{How to test your pipeline vs ours?}\label{sec:howto}
All FRIPON single-station observation files used in this study, along with a summary table of the triangulation results for each data reduction pipeline, are available on Zenodo\footnote{\url{https://doi.org/10.5281/zenodo.14963219}}. 

For access to pipelines compared in this paper, WMPL is open-source software\footnote{\url{https://github.com/wmpg/WesternMeteorPyLib}}. The DFN, FRIPON, and AMOS pipelines are currently being refactored to remove network-specific dependencies and improve documentation, with the aim of making them publicly available. In the interim, please contact the authors.

In this comparison, it is also vital to note that the per–station astrometry in the shared GFE files was only computed by the FRIPON pipeline. These GFE files, with FRIPON reduced astrometry, were then used by the DFN, WMPL, and AMOS. Our cross‑pipeline differences, therefore, reflect trajectory/velocity modelling and uncertainty propagation rather than astrometric/timing calibration.

\section{Conclusions}
By reprocessing 584 FRIPON‐observed Geminid fireballs with four independent data reduction pipelines (FRIPON, DFN, WMPL, and AMOS), we found that all achieve results similar to previous studies; however, with some differences in velocity and radiant coordinates. The per–station astrometric solutions produced by FRIPON were used by the DFN, WMPL, and AMOS; therefore, the cross‑pipeline differences we discuss do not arise from astrometric processing. FRIPON retrieved velocities around 0.2-0.4\,km\,s$^{-1}$ faster than DFN, WMPL, and AMOS. This overestimation is due to the deceleration-based model \citep{Bronshten1983} it uses to estimate velocities, which are not accurate when deceleration is minimal. The DFN, WMPL, and AMOS each employ more advanced velocity solvers, which produce similar results as long as the observations do not contain significant noise. Otherwise, when only 2 cameras have observed the fireball, or there are fewer than 50 data points, the DFN tends to estimate slightly lower velocities relative to WMPL and AMOS. Concerning the radiant, FRIPON solutions are systematically shifted by about $-0.3^{\circ}$ in the right ascension relative to the other pipelines due to an epoch misalignment (no final precession to J2000), and possibly compounded by FRIPON using an older ITRF93 model. This error is currently being remedied, and the FRIPON pipeline will, in the future, output correctly shifted J2000 radiants. All pipelines converge on orbital solutions that remain consistent with the known Geminid stream parameters, with over 99\% of fireballs retaining a $D_{N}<0.1$ relative to published Geminid observations. 

A more pressing discrepancy emerges in the formal velocity uncertainties. Even though each pipeline generally fits the raw observations to the same degree (i.e., the RMS of the spatial residuals has a similar magnitude), the reported 1-$\sigma$ velocity uncertainties differ systematically, with WMPL reporting the lowest velocity uncertainties, followed by AMOS, FRIPON, and then DFN. These inconsistencies primarily reflect contrasting treatments of astrometric weighting, timing precision, and systematic effects rather than deficiencies in any single solver. Reliable orbital uncertainties are crucial for modelling meteoroid streams and linking meteors to their parent bodies; achieving a consensus on uncertainty propagation is key for the community to address. Future work should focus on standardised approaches to specifying astrometric uncertainty, modelling pre-luminous deceleration, and automating outlier handling, thereby improving consistency and interpretability across diverse meteor networks. 

As meteor and fireball networks become more numerous due to the decreasing cost of precise sensors, along with the increased ubiquity of other observations from seismic, infrasound, radar, weather radar, telescopic, and satellite observations, we believe that this study represents the first step toward improving collaboration between different networks and data sources. By combining more sensors and integrating increasingly diverse data reduction methods, we anticipate that future studies will enable a more comprehensive understanding of meteoroid populations and their dynamics. In future work, we plan to conduct similar analyses for meteorite-dropping fireballs and pre-impact telescopic detections, such as the 2023 CX1 data, as having an external and independent estimate of the orbit would be extremely beneficial to the work.

\begin{acknowledgements}
This project received funding from the European Union’s Horizon 2020 research and innovation programme under the Marie Skłodowska-Curie grant agreement No945298 ParisRegionFP (P.M.S.), and the grant agreement No. 101150536 (S.A.)

The Global Fireball Observatory and data pipeline is enabled by the support of the Australian Research Council (DP230100301, LE170100106). 

FRIPON was initiated by funding from ANR (grant N.13-BS05-0009-03), carried by the Paris Observatory, Muséum National d’Histoire Naturelle, Paris-Saclay University and Institut Pythéas (LAM-CEREGE). VigieCiel was part of the 65 Millions d’Observateurs project, carried by the Muséum National d’Histoire Naturelle and funded by the French Investissements d’Avenir program. FRIPON data are hosted and processed at Institut Pythéas SIP (Service Informatique Pythéas), and a mirror is hosted at LTE (Le Laboratoire Temps Espace / Paris Observatory).

This project was partially funded by the Excellent Grant of Comenius University no. UK/3055/2024 (Flux of primitive and differentiated material to Earth). This work was also supported by the Slovak Grant Agency for Science (grant VEGA 1/0218/22), and by the Slovak Research and Development Agency grants APVV-23-0323 and VV-MVP-24-0232.

The authors would also like to thank Simon Jeanne for his communication and help in interpreting some of the results. Also, we would like to thank the anonymous reviewer whose thorough analysis and recommendations greatly improved the manuscript. 
\end{acknowledgements}

\bibliographystyle{aa}
\bibliography{references} 

\begin{thebibliography}{105}
\expandafter\ifx\csname natexlab\endcsname\relax\def\natexlab#1{#1}\fi

\bibitem[{Anderson {et~al.}(2022)Anderson, Towner, Fairweather, Bland,
  Devillepoix, Sansom, Cupak, Shober, \& Benedix}]{anderson2022successful}
Anderson, S.~L., Towner, M.~C., Fairweather, J., {et~al.} 2022, ApJL, 930, L25

\bibitem[{Anghel {et~al.}(2019)Anghel, Birlan, Nedelcu, \&
  Boaca}]{anghel2019photometric}
Anghel, S., Birlan, M., Nedelcu, D.-A., \& Boaca, I. 2019, RoAJ, 29, 191

\bibitem[{{Anghel} {et~al.}(2021){Anghel}, {Drolshagen}, {Ott}, {Birlan},
  {Colas}, {Nedelcu}, {Koschny}, {Zanda}, {Bouley}, {Jeanne}, \&
  et~al.}]{2021MNRAS.508.5716A}
{Anghel}, S., {Drolshagen}, E., {Ott}, T., {et~al.} 2021, \mnras, 508, 5716

\bibitem[{{Anghel} {et~al.}(2023){Anghel}, {Nedelcu}, {Birlan}, \&
  {Boaca}}]{2023MNRAS.518.2810A}
{Anghel}, S., {Nedelcu}, D.~A., {Birlan}, M., \& {Boaca}, I. 2023, MNRAS, 518,
  2810

\bibitem[{Antier(2023)}]{Antier_Par}
Antier, K. 2023, Bolide du 10 septembre, 00H13

\bibitem[{Arlt {et~al.}(2022)Arlt, Asher, Brown, Gural, Koschack, Molau, Rault,
  Rendtel, Richter, Verbeeck, \& et~al.}]{IMO_Handbook_2022}
Arlt, R., Asher, D.~J., Brown, P.~G., {et~al.} 2022, Handbook for Meteor
  Observers, ed. J.~Rendtel (International Meteor Organization)

\bibitem[{Barghini {et~al.}(2024)Barghini, Carbognani, Gardiol, Bertocco,
  Di~Carlo, Di~Martino, Falco, Morelli, Pratesi, Riva,
  {et~al.}}]{barghini2024recovery}
Barghini, D., Carbognani, A., Gardiol, D., {et~al.} 2024, in XIX Congresso
  Nazionale di Scienze Planetarie, Societ{\`a} Italiana di Scienze Planetarie,
  0--0

\bibitem[{Barghini {et~al.}(2025)Barghini, {\v{D}}uri{\v{s}}ov{\'a}, Koten,
  Bertaina, Gardiol, \& Hajdukov{\'a}}]{barghini2025kresaks}
Barghini, D., {\v{D}}uri{\v{s}}ov{\'a}, S., Koten, P., {et~al.} 2025, A\&A,
  701, A135

\bibitem[{Bessell(2000)}]{bessell2000hipparcos}
Bessell, M.~S. 2000, Publ. Astron. Soc. Pac., 112, 961

\bibitem[{Bland {et~al.}(2012)Bland, Spurn{\'y}, Bevan, Howard, Towner,
  Benedix, Greenwood, Shrben{\'y}, Franchi, Deacon,
  {et~al.}}]{bland2012australian}
Bland, P., Spurn{\'y}, P., Bevan, A., {et~al.} 2012, Aust. J. Earth Sci., 59,
  177

\bibitem[{Bland {et~al.}(2009)Bland, Spurn{\`y}, Towner, Bevan, Singleton,
  Bottke~Jr, Greenwood, Chesley, Shrben{\`y}, Borovi{\v{c}}ka,
  {et~al.}}]{bland2009anomalous}
Bland, P.~A., Spurn{\`y}, P., Towner, M.~C., {et~al.} 2009, Science, 325, 1525

\bibitem[{{Borovicka}(1990)}]{Borovicka_1990BAICz}
{Borovicka}, J. 1990, Bull. astr. Inst. Czechosl., 41, 391

\bibitem[{Borovi{\^c}ka \& Kalenda(2003)}]{borovicka2003moravka}
Borovi{\^c}ka, J. \& Kalenda, P. 2003, Meteorit. Planet. Sci., 38, 1023

\bibitem[{{Borovicka} {et~al.}(1995){Borovicka}, {Spurny}, \&
  {Keclikova}}]{1995A&AS..112..173B}
{Borovicka}, J., {Spurny}, P., \& {Keclikova}, J. 1995, \aaps, 112, 173

\bibitem[{Borovi{\v{c}}ka {et~al.}(2007)Borovi{\v{c}}ka, Spurn{\'y}, \&
  Koten}]{borovivcka2007atmospheric}
Borovi{\v{c}}ka, J., Spurn{\'y}, P., \& Koten, P. 2007, A\&A, 473, 661

\bibitem[{Borovi{\v{c}}ka {et~al.}(2020)Borovi{\v{c}}ka, Spurn{\`y}, \&
  Shrben{\`y}}]{borovivcka2020two}
Borovi{\v{c}}ka, J., Spurn{\`y}, P., \& Shrben{\`y}, L. 2020, AJ, 160, 42

\bibitem[{Borovi{\v{c}}ka {et~al.}(2022{\natexlab{a}})Borovi{\v{c}}ka,
  Spurn{\`y}, \& Shrben{\`y}}]{borovivcka2022_two}
Borovi{\v{c}}ka, J., Spurn{\`y}, P., \& Shrben{\`y}, L. 2022{\natexlab{a}},
  A\&A, 667, A158

\bibitem[{Borovi{\v{c}}ka {et~al.}(2022{\natexlab{b}})Borovi{\v{c}}ka,
  Spurn{\`y}, Shrben{\`y}, {\v{S}}tork, Kotkov{\'a}, Fuchs, Kecl{\'\i}kov{\'a},
  Zichov{\'a}, M{\'a}nek, V{\'a}chov{\'a}, {et~al.}}]{borovivcka2022_one}
Borovi{\v{c}}ka, J., Spurn{\`y}, P., Shrben{\`y}, L., {et~al.}
  2022{\natexlab{b}}, A\&A, 667, A157

\bibitem[{Borovi{\v{c}}ka {et~al.}(2013)Borovi{\v{c}}ka, T{\'o}th, Igaz,
  Spurn{\`y}, Kalenda, Haloda, Svore{\v{n}}, Korno{\v{s}}, Silber, Brown,
  {et~al.}}]{borovivcka2013kovsice}
Borovi{\v{c}}ka, J., T{\'o}th, J., Igaz, A., {et~al.} 2013, Meteorit. Planet.
  Sci., 48, 1757

\bibitem[{{Bronshtehn}(1983)}]{Bronshten1983}
{Bronshtehn}, V.~A. 1983, {Physics of meteor phenomena.}

\bibitem[{Brown {et~al.}(2010{\natexlab{a}})Brown, Weryk, Kohut, Edwards, \&
  Krzeminski}]{brown2010development}
Brown, P., Weryk, R., Kohut, S., Edwards, W., \& Krzeminski, Z.
  2010{\natexlab{a}}, WGN, vol. 38, no. 1, p. 25-30, 38, 25

\bibitem[{Brown {et~al.}(2008)Brown, Weryk, Wong, \&
  Jones}]{brown2008meteoroid}
Brown, P., Weryk, R., Wong, D., \& Jones, J. 2008, Icarus, 195, 317

\bibitem[{Brown {et~al.}(2010{\natexlab{b}})Brown, Wong, Weryk, \&
  Wiegert}]{brown2010meteoroid}
Brown, P., Wong, D., Weryk, R., \& Wiegert, P. 2010{\natexlab{b}}, Icarus, 207,
  66

\bibitem[{Brown {et~al.}(2023)Brown, McCausland, Hildebrand, Hanton, Eckart,
  Busemann, Krietsch, Maden, Welten, Caffee, {et~al.}}]{brown2023golden}
Brown, P.~G., McCausland, P., Hildebrand, A., {et~al.} 2023, Meteorit. Planet.
  Sci., 58, 1773

\bibitem[{Campbell-Burns \& Kacerek(2014)}]{campbell2014uk}
Campbell-Burns, P. \& Kacerek, R. 2014, WGN, 42, 139

\bibitem[{Ceplecha(1961)}]{ceplecha1961multiple}
Ceplecha, Z. 1961, Bull. astr. Inst. Czechosl., 12, 21

\bibitem[{{Ceplecha}(1987)}]{1987BAICz..38..222C}
{Ceplecha}, Z. 1987, Bull. astr. Inst. Czechosl., 38, 222

\bibitem[{Ceplecha {et~al.}(1998)Ceplecha, Borovi{\v{c}}ka, Elford, ReVelle,
  Hawkes, Porub{\v{c}}an, \& {\v{S}}imek}]{ceplecha1998meteor}
Ceplecha, Z., Borovi{\v{c}}ka, J., Elford, W.~G., {et~al.} 1998, Space Sci.
  Rev., 84, 327

\bibitem[{Ceplecha {et~al.}(1993)Ceplecha, Spurny, Borovicka, \&
  Keclikova}]{ceplecha1993atmospheric}
Ceplecha, Z., Spurny, P., Borovicka, J., \& Keclikova, J. 1993, A\&A, 279, 615

\bibitem[{Colas {et~al.}(2020)Colas, Zanda, Bouley, Jeanne, Malgoyre, Birlan,
  Blanpain, Gattacceca, Jorda, Lecubin, {et~al.}}]{colas2020fripon}
Colas, F., Zanda, B., Bouley, S., {et~al.} 2020, A\&A, 644, A53

\bibitem[{Cooke \& Moser(2012)}]{cooke2012status}
Cooke, W.~J. \& Moser, D.~E. 2012, in Proceedings of the International Meteor
  Conference, 30th IMC, Sibiu, Romania, 2011, 9--12

\bibitem[{Devillepoix {et~al.}(2020)Devillepoix, Cupak, Bland, Sansom, Towner,
  Howie, Hartig, Jansen-Sturgeon, Shober, Anderson,
  {et~al.}}]{devillepoix2020global}
Devillepoix, H., Cupak, M., Bland, P., {et~al.} 2020, Planet. Space Sci., 191,
  105036

\bibitem[{Devillepoix {et~al.}(2019)Devillepoix, Bland, Sansom, Towner, Cupák,
  Howie, Hartig, Jansen-Sturgeon, \& Cox}]{devillepoix2019observation}
Devillepoix, H.~A., Bland, P.~A., Sansom, E.~K., {et~al.} 2019, MNRAS, 483,
  5166

\bibitem[{Devillepoix {et~al.}(2018)Devillepoix, Sansom, Bland, Towner,
  Cup{\'a}k, Howie, Jansen-Sturgeon, Cox, Hartig, Benedix,
  {et~al.}}]{devillepoix2018dingle}
Devillepoix, H.~A., Sansom, E.~K., Bland, P.~A., {et~al.} 2018, Meteorit.
  Planet. Sci., 53, 2212

\bibitem[{{Devillepoix} {et~al.}(2022){Devillepoix}, {Sansom}, {Shober},
  {Anderson}, {Towner}, {Lagain}, {Cup{\'a}k}, {Bland}, {Howie},
  {Jansen-Sturgeon}, {Hartig}, {Sokolowski}, {Benedix}, \&
  {Forman}}]{Devillepoix_Madura_Cave}
{Devillepoix}, H. A.~R., {Sansom}, E.~K., {Shober}, P., {et~al.} 2022,
  Meteorit. Planet. Sci., 57, 1328

\bibitem[{Drummond(1981)}]{drummond1981test}
Drummond, J.~D. 1981, Icarus, 45, 545

\bibitem[{{Duris} {et~al.}(2018){Duris}, {Kornos}, \& {Toth}}]{Duris2018}
{Duris}, F., {Kornos}, L., \& {Toth}, J. 2018, in Proceedings of the
  International Meteor Conference, 37th IMC, Pezinok-Modra, Slovakia, 2018, ed.
  R.~{Rudawska}, J.~{Rendtel}, C.~{Powell}, R.~{Lunsford}, C.~{Verbeeck}, \&
  A.~{Knofel}, 127--128

\bibitem[{Dyl {et~al.}(2016)Dyl, Benedix, Bland, Friedrich, Spurn{\`y}, Towner,
  O'Keefe, Howard, Greenwood, Macke, {et~al.}}]{dyl2016characterization}
Dyl, K.~A., Benedix, G.~K., Bland, P.~A., {et~al.} 2016, Meteorit. Planet.
  Sci., 51, 596

\bibitem[{Egal {et~al.}(2017)Egal, Gural, Vaubaillon, Colas, \&
  Thuillot}]{egal2017challenge}
Egal, A., Gural, P., Vaubaillon, J., Colas, F., \& Thuillot, W. 2017, Icarus,
  294, 43

\bibitem[{{Egal} {et~al.}(2025){Egal}, {Vida}, {Colas}, {Zanda}, {Bouley},
  {Steinhausser}, {Vernazza}, {Ferri{\`e}re}, {Gattacceca}, {Birlan},
  {Vaubaillon}, {Antier}, {Anghel}, {Desmars}, {Bailli{\'e}}, {Maquet},
  {Bouquillon}, {Malgoyre}, {Jeanne}, {Borovi{\v{c}}ka}, {Spurn{\'y}},
  {Devillepoix}, {Micheli}, {Farnocchia}, {Naidu}, {Brown}, {Wiegert},
  {S{\'a}rneczky}, {P{\'a}l}, {Moskovitz}, {Kareta}, {Santana-Ros}, {Le
  Pichon}, {Mazet-Roux}, {Vergoz}, {McFadden}, {Assink}, {Evers}, {Krietsch},
  {Busemann}, {Maden}, {Eckart}, {Barrat}, {Povinec}, {Sykora}, {Kontul'},
  {Marchhart}, {Martschini}, {Merchel}, {Wieser}, {Gounelle}, {Pont},
  {Sans-Jofre}, {de Vet}, {Baziotis}, {Bro{\v{z}}}, {Marsset}, {Vergne},
  {Hanu{\v{s}}}, {Devog{\`e}le}, {Conversi}, {Oca{\~n}a}, {Buzzi}, {Alin
  Nedelcu}, {Sonka}, {Losse}, {Dupouy}, {Korlevi{\'c}}, {Husar}, {Jahn},
  {{\v{S}}egon}, {McIntyre}, {Neubert}, {Beck}, {Shober}, {Lagain},
  {Trigo-Rodriguez}, {Herrero}, {Rowe}, {Smedley}, {King}, {Sylla}, {Gardiol},
  {Barghini}, {Lamy}, {Jehin}, {Koschny}, {Poppe}, {Jord{\'a}n}, {Mendez},
  {Vieira}, {Cremades}, {Chennaoui Aoudjehane}, {Benkhaldoun}, {Hernandez},
  {Robertson}, \& {Jenniskens}}]{Egal2025_SPLV}
{Egal}, A., {Vida}, D., {Colas}, F., {et~al.} 2025, Nat. Astron.

\bibitem[{Gardiol {et~al.}(2021)Gardiol, Barghini, Buzzoni, Carbognani,
  Di~Carlo, Di~Martino, Knapic, Londero, Pratesi, Rasetti,
  {et~al.}}]{gardiol2021cavezzo}
Gardiol, D., Barghini, D., Buzzoni, A., {et~al.} 2021, MNRAS, 501, 1215

\bibitem[{Granvik {et~al.}(2018)Granvik, Morbidelli, Jedicke, Bolin, Bottke,
  Beshore, Vokrouhlick{\'y}, Nesvorn{\'y}, \& Michel}]{granvik2018debiased}
Granvik, M., Morbidelli, A., Jedicke, R., {et~al.} 2018, Icarus, 312, 181

\bibitem[{Gritsevich(2009)}]{gritsevich2009determination}
Gritsevich, M. 2009, Adv. Space Res., 44, 323

\bibitem[{Gural \& Segon(2009)}]{gural2009new}
Gural, P. \& Segon, D. 2009, WGN, vol. 37, no. 1, p. 28-32, 37, 28

\bibitem[{Gural(2011)}]{gural2011california}
Gural, P.~S. 2011, in Proceedings of the International Meteor Conference, 29th
  IMC, Armagh, Northern Ireland, 2010, 28--31

\bibitem[{Gural(2012)}]{gural2012new}
Gural, P.~S. 2012, Meteorit. Planet. Sci., 47, 1405

\bibitem[{Hajdukova {et~al.}(2024)Hajdukova, Stober, Barghini, Koten,
  Vaubaillon, Sterken, {\v{D}}uri{\v{s}}ov{\'a}, Jackson, \&
  Desch}]{hajdukova2024no}
Hajdukova, M., Stober, G., Barghini, D., {et~al.} 2024, A\&A, 691, A8

\bibitem[{Hajdukov{\'a}~Jr {et~al.}(2017)Hajdukov{\'a}~Jr, Koten, Korno{\v{s}},
  \& T{\'o}th}]{hajdukova2017meteoroid}
Hajdukov{\'a}~Jr, M., Koten, P., Korno{\v{s}}, L., \& T{\'o}th, J. 2017,
  Planet. Space Sci., 143, 89

\bibitem[{Halliday {et~al.}(1978)Halliday, Blackwell, \&
  Griffin}]{halliday1978innisfree}
Halliday, I., Blackwell, A.~T., \& Griffin, A.~A. 1978, JRASC, 72, 15

\bibitem[{Hankey {et~al.}(2020)Hankey, Perlerin, \& Meisel}]{hankey2020all}
Hankey, M., Perlerin, V., \& Meisel, D. 2020, Planet. Space Sci., 190, 105005

\bibitem[{Henych {et~al.}(2024)Henych, Borovi{\v{c}}ka, Voj{\'a}{\v{c}}ek, \&
  Spurn{\`y}}]{henych2024mechanical}
Henych, T., Borovi{\v{c}}ka, J., Voj{\'a}{\v{c}}ek, V., \& Spurn{\`y}, P. 2024,
  A\&A, 683, A229

\bibitem[{Howie {et~al.}(2017{\natexlab{a}})Howie, Paxman, Bland, Towner,
  Cupák, Sansom, \& Devillepoix}]{howie2017build}
Howie, R.~M., Paxman, J., Bland, P.~A., {et~al.} 2017{\natexlab{a}}, Exp.
  Astron., 43, 237

\bibitem[{Howie {et~al.}(2017{\natexlab{b}})Howie, Paxman, Bland, Towner,
  Sansom, \& Devillepoix}]{howie2017submillisecond}
Howie, R.~M., Paxman, J., Bland, P.~A., {et~al.} 2017{\natexlab{b}}, Meteorit.
  Planet. Sci., 52, 1669

\bibitem[{Jansen-Sturgeon {et~al.}(2020)Jansen-Sturgeon, Sansom, Devillepoix,
  Bland, Towner, Howie, \& Hartig}]{jansen2020dynamic}
Jansen-Sturgeon, T., Sansom, E.~K., Devillepoix, H.~A., {et~al.} 2020, AJ, 160,
  190

\bibitem[{Jeanne(2020)}]{jeanne2020méthode}
Jeanne, S. 2020, PhD thesis, Universit{\'e} Paris Sciences et Lettres

\bibitem[{Jeanne {et~al.}(2019)Jeanne, Colas, Zanda, Birlan, Vaubaillon,
  Bouley, Vernazza, Jorda, Gattacceca, Rault, {et~al.}}]{jeanne2019calibration}
Jeanne, S., Colas, F., Zanda, B., {et~al.} 2019, A\&A, 627, A78

\bibitem[{Jenniskens(2008)}]{jenniskens2008meteoroid}
Jenniskens, P. 2008, Icarus, 194, 13

\bibitem[{Jenniskens {et~al.}(2011)Jenniskens, Gural, Dynneson, Grigsby,
  Newman, Borden, Koop, \& Holman}]{jenniskens2011cams}
Jenniskens, P., Gural, P., Dynneson, L., {et~al.} 2011, Icarus, 216, 40

\bibitem[{Jenniskens {et~al.}(2016)Jenniskens, N{\'e}non, Albers, Gural,
  Haberman, Holman, Morales, Grigsby, Samuels, \&
  Johannink}]{jenniskens2016established}
Jenniskens, P., N{\'e}non, Q., Albers, J., {et~al.} 2016, Icarus, 266, 331

\bibitem[{Jenniskens {et~al.}(2019)Jenniskens, Utas, Yin, Matson, Fries,
  Howell, Free, Albers, Devillepoix, Bland, {et~al.}}]{jenniskens2019creston}
Jenniskens, P., Utas, J., Yin, Q.-Z., {et~al.} 2019, Meteorit. Planet. Sci.,
  54, 699

\bibitem[{Jopek {et~al.}(2003)Jopek, Valsecchi, \&
  Froeschl{\'e}}]{jopek2003meteor}
Jopek, T., Valsecchi, G., \& Froeschl{\'e}, C. 2003, MNRAS, 344, 665

\bibitem[{Jopek(1993)}]{jopek1993remarks}
Jopek, T.~J. 1993, Icarus, 106, 603

\bibitem[{Jopek {et~al.}(2008)Jopek, Rudawska, \&
  Bartczak}]{jopek2008meteoroid}
Jopek, T.~J., Rudawska, R., \& Bartczak, P. 2008, Advances in Meteoroid and
  Meteor Science, 73

\bibitem[{{King} {et~al.}(2022){King}, {Daly}, {Rowe}, {Joy}, {Greenwood},
  {Devillepoix}, {Suttle}, {Chan}, {Russell}, {Bates}, {Bryson}, {Clay},
  {Vida}, {Lee}, {O'Brien}, {Hallis}, {Stephen}, {Tart{\`e}se}, {Sansom},
  {Towner}, {Cupak}, {Shober}, {Bland}, {Findlay}, {Franchi}, {Verchovsky},
  {Abernethy}, {Grady}, {Floyd}, {Van Ginneken}, {Bridges}, {Hicks}, {Jones},
  {Mitchell}, {Genge}, {Jenkins}, {Martin}, {Sephton}, {Watson}, {Salge},
  {Shirley}, {Curtis}, {Warren}, {Bowles}, {Stuart}, {Di Nicola}, {Gy{\"o}re},
  {Boyce}, {Shaw}, {Elliott}, {Steele}, {Povinec}, {Laubenstein}, {Sanderson},
  {Cresswell}, {Jull}, {S{\'y}kora}, {Sridhar}, {Harrison}, {Willcocks},
  {Harrison}, {Hallatt}, {Wozniakiewicz}, {Burchell}, {Alesbrook}, {Dignam},
  {Almeida}, {Smith}, {Clark}, {Humphreys-Williams}, {Schofield}, {Cornwell},
  {Spathis}, {Morgan}, {Perkins}, {Kacerek}, {Campbell-Burns}, {Colas},
  {Zanda}, {Vernazza}, {Bouley}, {Jeanne}, {Hankey}, {Collins}, {Young},
  {Shaw}, {Horak}, {Jones}, {James}, {Bosley}, {Shuttleworth}, {Dickinson},
  {McMullan}, {Robson}, {Smedley}, {Stanley}, {Bassom}, {McIntyre}, {Suttle},
  {Fleet}, {Bastiaens}, {Ih{\'a}sz}, {McMullan}, {Boazman}, {Dickeson},
  {Grindrod}, {Pickersgill}, {Weir}, {Suttle}, {Farrelly}, {Spencer}, {Naqvi},
  {Mayne}, {Skilton}, {Kirk}, {Mounsey}, {Mounsey}, {Mounsey}, {Godfrey},
  {Bond}, {Bond}, {Wilcock}, {Wilcock}, \& {Wilcock}}]{King_Winchcombe2022SciA}
{King}, A.~J., {Daly}, L., {Rowe}, J., {et~al.} 2022, Sci. Adv., 8, eabq3925

\bibitem[{Koten {et~al.}(2019)Koten, Rendtel, Shrben\'{y}, Gural, Borovicka, \&
  Kozak}]{koten2019meteors}
Koten, P., Rendtel, J., Shrben\'{y}, L., {et~al.} 2019, Meteoroids: Sources of
  Meteors on Earth and Beyond, 90

\bibitem[{Kres{\'a}k \& Porubcan(1970)}]{kresak1970dispersion}
Kres{\'a}k, L. \& Porubcan, V. 1970, Bull. astr. Inst. Czechosl., 21, 153

\bibitem[{Lyytinen \& Gritsevich(2016)}]{lyytinen2016implications}
Lyytinen, E. \& Gritsevich, M. 2016, Planet. Space Sci., 120, 35

\bibitem[{Matlovi{\v{c}} {et~al.}(2019)Matlovi{\v{c}}, T{\'o}th, Rudawska,
  Korno{\v{s}}, \& Pisar{\v{c}}{\'\i}kov{\'a}}]{matlovic2019}
Matlovi{\v{c}}, P., T{\'o}th, J., Rudawska, R., Korno{\v{s}}, L., \&
  Pisar{\v{c}}{\'\i}kov{\'a}, A. 2019, A\&A, 629, A71

\bibitem[{{Matlovi{\v{c}}} {et~al.}(2020){Matlovi{\v{c}}}, {Korno{\v{s}}},
  {Kov{\'a}{\v{c}}ov{\'a}}, {T{\'o}th}, \& {Licandro}}]{matlovic2020}
{Matlovi{\v{c}}}, P., {Korno{\v{s}}}, L., {Kov{\'a}{\v{c}}ov{\'a}}, M.,
  {T{\'o}th}, J., \& {Licandro}, J. 2020, \aap, 636, A122

\bibitem[{{Matlovi{\v{c}}} {et~al.}(2022){Matlovi{\v{c}}},
  {Pisar{\v{c}}{\'\i}kov{\'a}}, {T{\'o}th}, {Mach}, {{\v{C}}erm{\'a}k},
  {Loehle}, {Korno{\v{s}}}, {Ferri{\`e}re}, {{\v{S}}ilha}, {Leiser}, \&
  {Ravichandran}}]{2022MNRAS.513.3982M}
{Matlovi{\v{c}}}, P., {Pisar{\v{c}}{\'\i}kov{\'a}}, A., {T{\'o}th}, J.,
  {et~al.} 2022, \mnras, 513, 3982

\bibitem[{McCrosky {et~al.}(1971)McCrosky, Posen, Schwartz, \&
  Shao}]{mccrosky1971lost}
McCrosky, R.~E., Posen, A., Schwartz, G., \& Shao, C.-Y. 1971, J. Geophys.
  Res., 76, 4090

\bibitem[{McMullan {et~al.}(2023)McMullan, Vida, Devillepoix, Rowe, Daly, King,
  Cup{\'a}k, Howie, Sansom, Shober, {et~al.}}]{mcmullan2023winchcombe}
McMullan, S., Vida, D., Devillepoix, H.~A., {et~al.} 2023, Meteorit. Planet.
  Sci.

\bibitem[{Molau(1999)}]{molau1999meteor}
Molau, S. 1999, in Proceedings of the International Meteor Conference, 17th
  IMC, Stara Lesna, Slovakia, 1998, 9--16

\bibitem[{Molau(2001)}]{molau2001akm}
Molau, S. 2001, in Meteoroids 2001 Conference, Vol. 495, 315--318

\bibitem[{Moorhead {et~al.}(2021)Moorhead, Clements, \&
  Vida}]{moorhead2021meteor}
Moorhead, A.~V., Clements, T., \& Vida, D. 2021, MNRAS, 508, 326

\bibitem[{Pe{\~n}a-Asensio {et~al.}(2021)Pe{\~n}a-Asensio,
  Trigo-Rodr{\'\i}guez, Gritsevich, \& Rimola}]{pena2021accurate}
Pe{\~n}a-Asensio, E., Trigo-Rodr{\'\i}guez, J.~M., Gritsevich, M., \& Rimola,
  A. 2021, MNRAS, 504, 4829

\bibitem[{Rowe(2021)}]{rowe2021just}
Rowe, J. 2021, WGN, 49, 211

\bibitem[{Sansom {et~al.}(2015)Sansom, Bland, Paxman, \&
  Towner}]{sansom2015novel}
Sansom, E.~K., Bland, P., Paxman, J., \& Towner, M. 2015, Meteorit. Planet.
  Sci., 50, 1423

\bibitem[{Sansom {et~al.}(2020)Sansom, Bland, Towner, Devillepoix, Cup{\'A}k,
  Howie, Jansen-Sturgeon, Cox, Hartig, Paxman, {et~al.}}]{sansom2020murrili}
Sansom, E.~K., Bland, P.~A., Towner, M.~C., {et~al.} 2020, Meteorit. Planet.
  Sci., 55, 2157

\bibitem[{Sansom {et~al.}(2019{\natexlab{a}})Sansom, Gritsevich, Devillepoix,
  Jansen-Sturgeon, Shober, Bland, Towner, Cupák, Howie, \&
  Hartig}]{sansom2019determining}
Sansom, E.~K., Gritsevich, M., Devillepoix, H.~A., {et~al.} 2019{\natexlab{a}},
  ApJ, 885, 115

\bibitem[{Sansom {et~al.}(2019{\natexlab{b}})Sansom, Jansen-Sturgeon, Rutten,
  Devillepoix, Bland, Howie, Cox, Towner, Cupák, \& Hartig}]{sansom20193d}
Sansom, E.~K., Jansen-Sturgeon, T., Rutten, M.~G., {et~al.} 2019{\natexlab{b}},
  Icarus, 321, 388

\bibitem[{Shober {et~al.}(2022)Shober, Devillepoix, Sansom, Towner, Cup{\'a}k,
  Anderson, Benedix, Forman, Bland, Howie, {et~al.}}]{shober2022arpu}
Shober, P.~M., Devillepoix, H.~A., Sansom, E.~K., {et~al.} 2022, Meteorit.
  Planet. Sci., 57, 1146

\bibitem[{Shober {et~al.}(2025)Shober, Devillepoix, Vaubaillon, Anghel, Deam,
  Sansom, Colas, Zanda, Vernazza, \& Bland}]{shober2025perihelion}
Shober, P.~M., Devillepoix, H.~A., Vaubaillon, J., {et~al.} 2025, Nat. Astron.,
  1

\bibitem[{Shober {et~al.}(2020)Shober, Jansen-Sturgeon, Sansom, Devillepoix,
  Towner, Bland, Cupák, Howie, \& Hartig}]{shober2020did}
Shober, P.~M., Jansen-Sturgeon, T., Sansom, E.~K., {et~al.} 2020, AJ, 159, 191

\bibitem[{Shober {et~al.}(2021)Shober, Sansom, Bland, Devillepoix, Towner,
  Cup{\'a}k, Howie, Hartig, \& Anderson}]{shober2021main}
Shober, P.~M., Sansom, E.~K., Bland, P.~A., {et~al.} 2021, PSJ, 2, 98

\bibitem[{Shober \& Vaubaillon(2024)}]{shober2024generalizable}
Shober, P.~M. \& Vaubaillon, J. 2024, A\&A, 686, A130

\bibitem[{SonotaCo(2009)}]{sonotaco2009meteor}
SonotaCo, A. 2009, WGN, 37, 55

\bibitem[{Southworth \& Hawkins(1963)}]{southworth1963statistics}
Southworth, R. \& Hawkins, G. 1963, Smithsonian Contr. Astrophys., 7, 261

\bibitem[{Spurn{\`y} {et~al.}(2006)Spurn{\`y}, Borovi{\v{c}}ka, \&
  Shrben{\`y}}]{spurny2006automation}
Spurn{\`y}, P., Borovi{\v{c}}ka, J., \& Shrben{\`y}, L. 2006, Proc. Int.
  Astron. Union, 2, 121

\bibitem[{Stulov {et~al.}(1995)Stulov, Mirskii, \&
  Vislyi}]{stulov1995aerodynamics}
Stulov, V., Mirskii, V., \& Vislyi, A. 1995, Moscow: Science. Fizmatlit

\bibitem[{Suk \& {\v{S}}imberov{\'a}(2017)}]{suk2017automated}
Suk, T. \& {\v{S}}imberov{\'a}, S. 2017, Earth Moon Planets, 120, 189

\bibitem[{T{\'o}th {et~al.}(2015)T{\'o}th, Korno{\v{s}}, Zigo, Gajdo{\v{s}},
  Kalman{\v{c}}ok, Vil{\'a}gi, {\v{S}}imon, Vere{\v{s}}, {\v{S}}ilha,
  Bu{\v{c}}ek, {et~al.}}]{toth2015all}
T{\'o}th, J., Korno{\v{s}}, L., Zigo, P., {et~al.} 2015, Planet. Space Sci.,
  118, 102

\bibitem[{{T{\'o}th} {et~al.}(2019){T{\'o}th}, {{\v{S}}ilha}, {Matlovi{\v{c}}},
  {Korno{\v{s}}}, {Zigo}, {Vil{\'a}gi}, {Kalman{\v{c}}ok}, {{\v{S}}imon}, \&
  {Vere{\v{s}}}}]{toth2019}
{T{\'o}th}, J., {{\v{S}}ilha}, J., {Matlovi{\v{c}}}, P., {et~al.} 2019, in 1st
  NEO and Debris Detection Conference\_ESA2019, 63

\bibitem[{Trigo-Rodr{\'\i}guez {et~al.}(2006)Trigo-Rodr{\'\i}guez, Llorca,
  Castro-Tirado, Ortiz, Docobo, \& Fabregat}]{trigo2006spanish}
Trigo-Rodr{\'\i}guez, J.~M., Llorca, J., Castro-Tirado, A.~J., {et~al.} 2006,
  A\&G, 47, 6

\bibitem[{Turchak \& Gritsevich(2014)}]{turchak2014meteoroids}
Turchak, L.~I. \& Gritsevich, M.~I. 2014, J. Theor. Appl. Mech., 44, 15

\bibitem[{Valsecchi {et~al.}(1999)Valsecchi, Jopek, \&
  Froeschl{\'e}}]{valsecchi1999meteoroid}
Valsecchi, G., Jopek, T., \& Froeschl{\'e}, C. 1999, MNRAS, 304, 743

\bibitem[{Vida {et~al.}(2018)Vida, Brown, \&
  Campbell-Brown}]{vida2018modelling}
Vida, D., Brown, P.~G., \& Campbell-Brown, M. 2018, MNRAS, 479, 4307

\bibitem[{Vida {et~al.}(2024)Vida, Brown, Campbell-Brown, \&
  Egal}]{vida2024first}
Vida, D., Brown, P.~G., Campbell-Brown, M., \& Egal, A. 2024, Icarus, 408,
  115842

\bibitem[{Vida {et~al.}(2023)Vida, Brown, Devillepoix, Wiegert, Moser,
  Matlovi{\v{c}}, Herd, Hill, Sansom, Towner, {et~al.}}]{vida2023direct}
Vida, D., Brown, P.~G., Devillepoix, H.~A., {et~al.} 2023, Nat. Astron., 7, 318

\bibitem[{Vida {et~al.}(2020{\natexlab{a}})Vida, Campbell-Brown, Brown, Egal,
  \& Mazur}]{vida2020new}
Vida, D., Campbell-Brown, M., Brown, P.~G., Egal, A., \& Mazur, M.~J.
  2020{\natexlab{a}}, A\&A, 635, A153

\bibitem[{Vida {et~al.}(2020{\natexlab{b}})Vida, Gural, Brown, Campbell-Brown,
  \& Wiegert}]{vida2020estimating}
Vida, D., Gural, P.~S., Brown, P.~G., Campbell-Brown, M., \& Wiegert, P.
  2020{\natexlab{b}}, MNRAS, 491, 2688

\bibitem[{Vida {et~al.}(2021)Vida, {\v{S}}egon, Gural, Brown, McIntyre,
  Dijkema, Pavleti{\'c}, Kuki{\'c}, Mazur, Eschman, {et~al.}}]{vida2021global}
Vida, D., {\v{S}}egon, D., Gural, P.~S., {et~al.} 2021, MNRAS, 506, 5046

\bibitem[{Weryk {et~al.}(2013)Weryk, Campbell-Brown, Wiegert, Brown,
  Krzeminski, \& Musci}]{weryk2013canadian}
Weryk, R., Campbell-Brown, M., Wiegert, P., {et~al.} 2013, Icarus, 225, 614

\bibitem[{Wi{\'s}niewski {et~al.}(2017)Wi{\'s}niewski, {\.Z}o{\l}{\k{a}}dek,
  Olech, Tyminski, Maciejewski, Fietkiewicz, Rudawska, Gozdalski,
  Gawro{\'n}ski, Suchodolski, {et~al.}}]{wisniewski2017current}
Wi{\'s}niewski, M., {\.Z}o{\l}{\k{a}}dek, P., Olech, A., {et~al.} 2017, Planet.
  Space Sci., 143, 12

\bibitem[{Zanda {et~al.}(2023)Zanda, {\'E}gal, Steinhausser, Bouley, Colas,
  Ferri{\`e}re, Vida, Devillepoix, Ma-Quet, Antier,
  {et~al.}}]{zanda2023rrecovery}
Zanda, B., {\'E}gal, A., Steinhausser, A., {et~al.} 2023, in 86th Annual
  Meeting of the Meteoritical Society

\end{thebibliography}

\clearpage
\onecolumn
\begin{appendix}           
\section{Pipeline-comparison matrix plots} \label{sec:appendix}

\begin{figure}[!ht]         
  \centering
  \includegraphics[width=\textwidth]{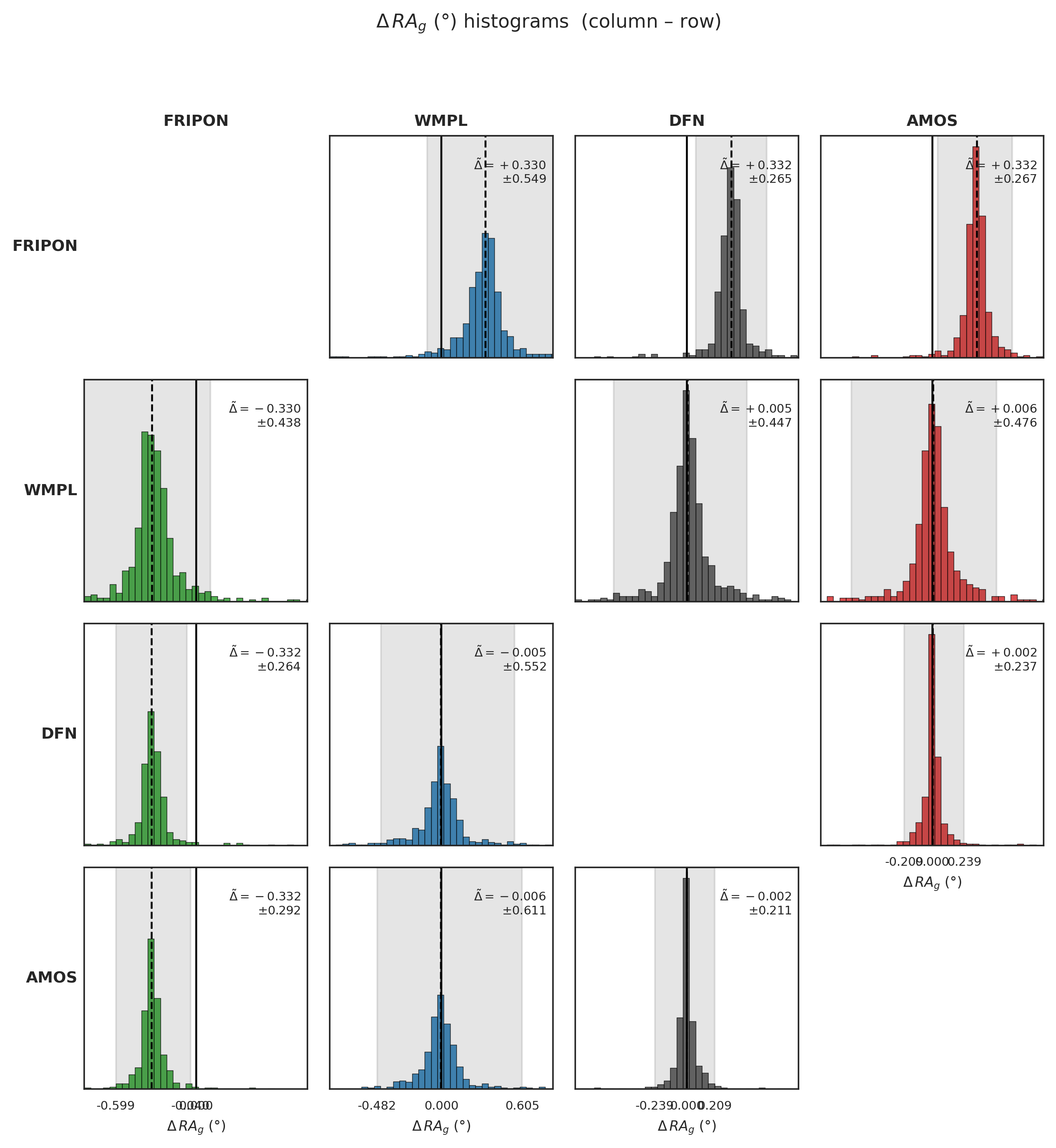}
  \caption{%
    $\Delta RA$ histograms (column – row) for all pairwise combinations of FRIPON, WMPL, DFN, and AMOS. The solid vertical line denotes zero, the dashed vertical line indicates the median, and the grey region encompasses the 95\% confidence region. $\Delta$ is the median value, and the $\pm$ indicates half of the 95\% region.
  }
  \label{fig:app_dRA}
\end{figure}

\begin{figure*}[p]         
  \centering
  \includegraphics[width=\textwidth]{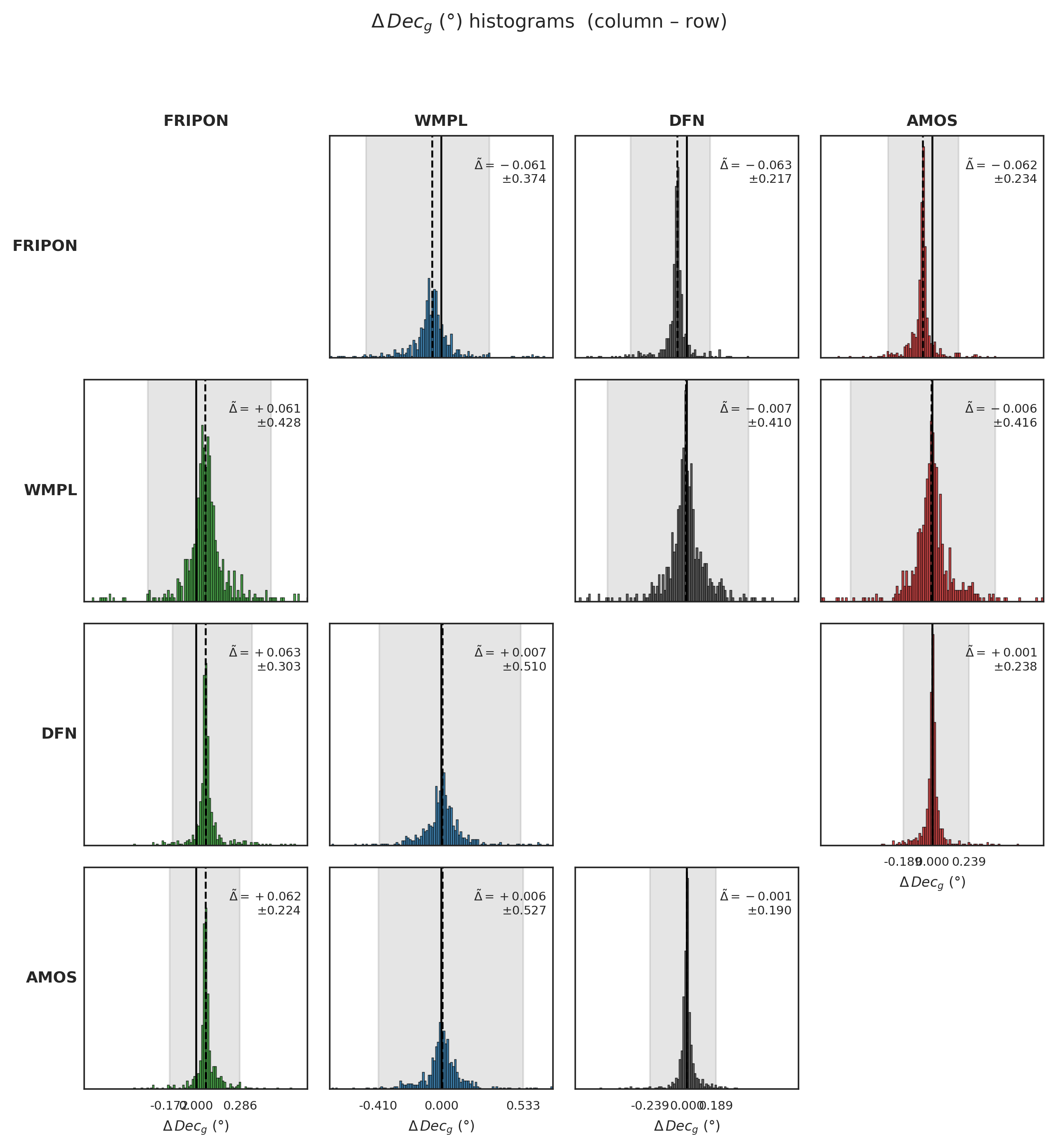}
  \caption{%
    $\Delta Dec$ histograms (column – row) for all pairwise combinations of FRIPON, WMPL, DFN, and AMOS. The solid vertical line denotes zero, the dashed vertical line indicates the median, and the grey region encompasses the 95\% confidence region. $\Delta$ is the median value, and the $\pm$ indicates half of the 95\% region.
  }
  \label{fig:app_dDec}
\end{figure*}

\begin{figure*}[p]         
  \centering
  \includegraphics[width=\textwidth]{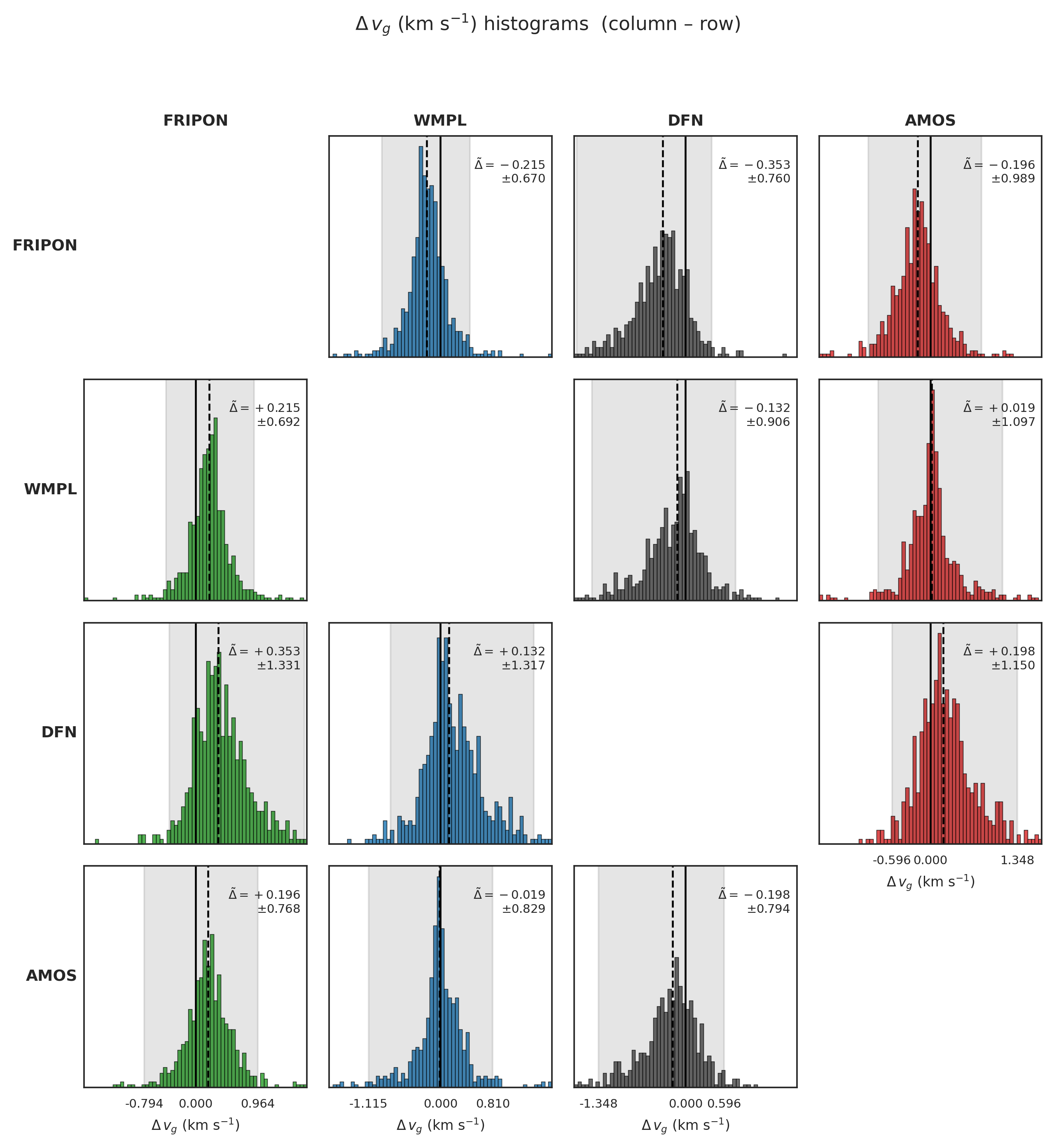}
  \caption{%
    $\Delta v_{g}$ histograms (column – row) for all pairwise combinations of FRIPON, WMPL, DFN, and AMOS. The solid vertical line denotes zero, the dashed vertical line indicates the median, and the grey region encompasses the 95\% confidence region. $\Delta$ is the median value, and the $\pm$ indicates half of the 95\% region.
  }
  \label{fig:app_dvg}
\end{figure*}

\begin{figure*}[p]         
  \centering
  \includegraphics[width=\textwidth]{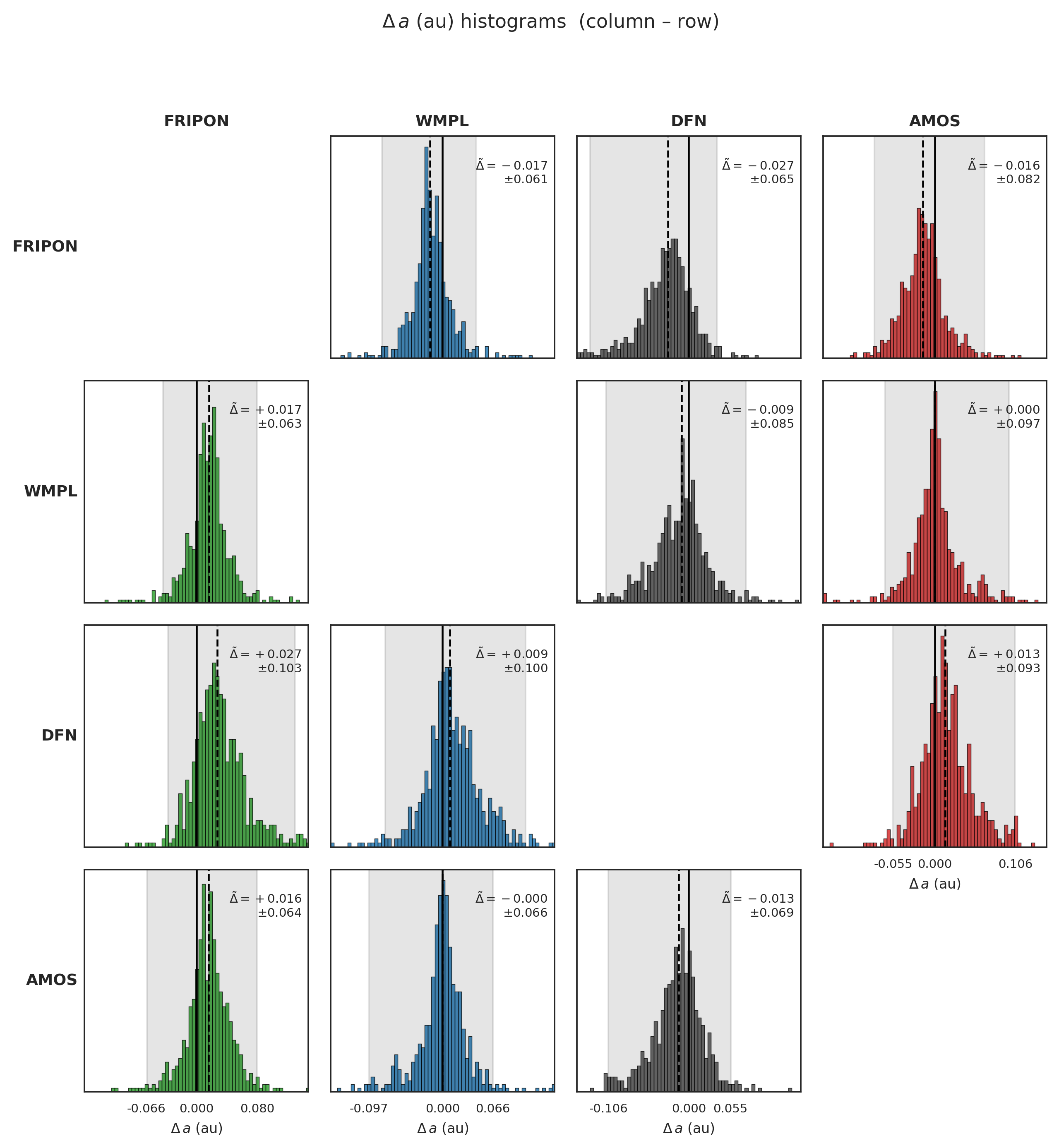}
  \caption{%
    $\Delta a$ histograms (column – row) for all pairwise combinations of FRIPON, WMPL, DFN, and AMOS. The solid vertical line denotes zero, the dashed vertical line indicates the median, and the grey region encompasses the 95\% confidence region. $\Delta$ is the median value, and the $\pm$ indicates half of the 95\% region.
  }
  \label{fig:app_da}
\end{figure*}

\begin{figure*}[p]        
  \centering
  \includegraphics[width=\textwidth]{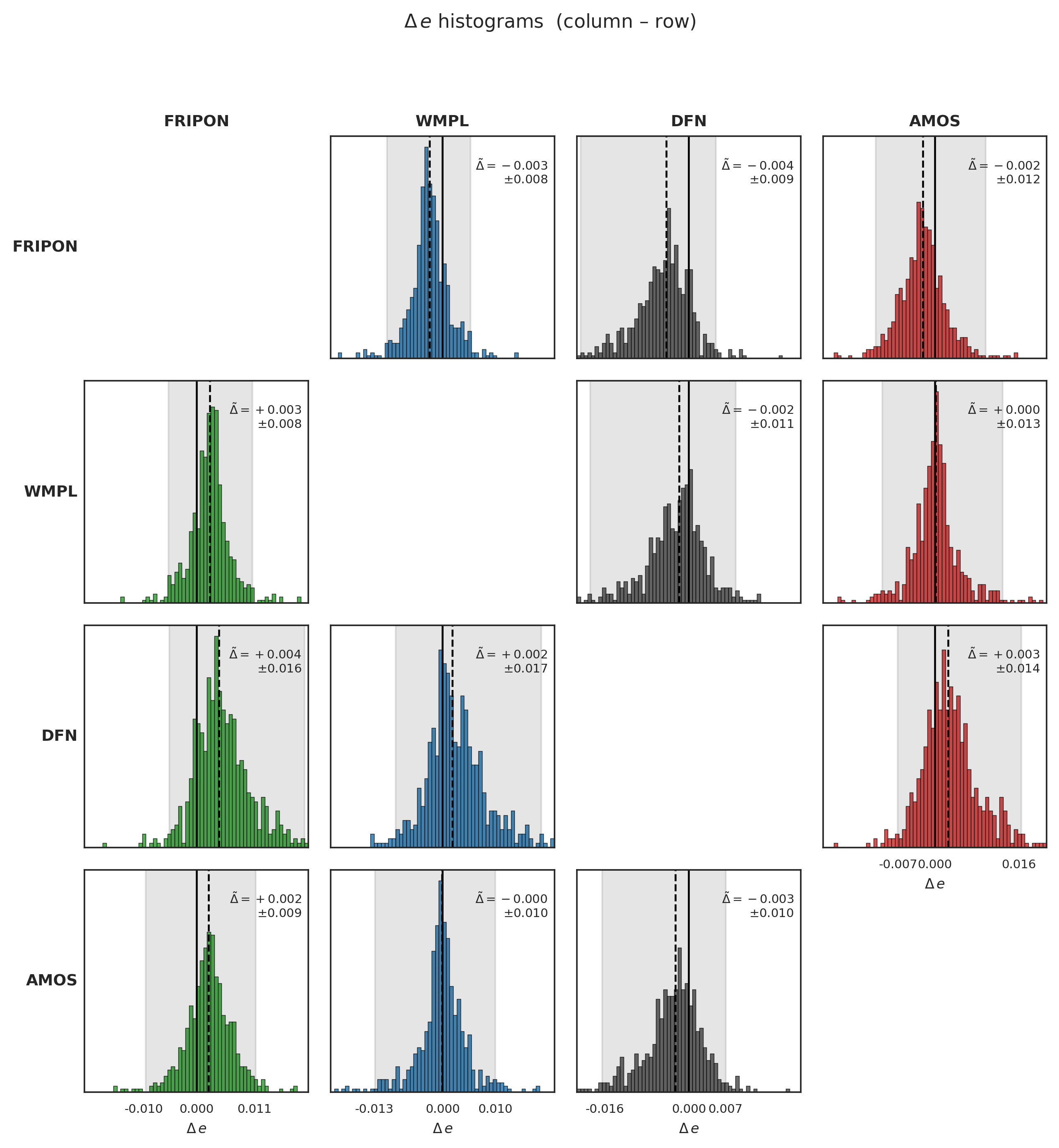}
  \caption{%
    $\Delta e$ histograms (column – row) for all pairwise combinations of FRIPON, WMPL, DFN, and AMOS. The solid vertical line denotes zero, the dashed vertical line indicates the median, and the grey region encompasses the 95\% confidence region. $\Delta$ is the median value, and the $\pm$ indicates half of the 95\% region.
  }
  \label{fig:app_de}
\end{figure*}

\begin{figure*}[p]        
  \centering
  \includegraphics[width=\textwidth]{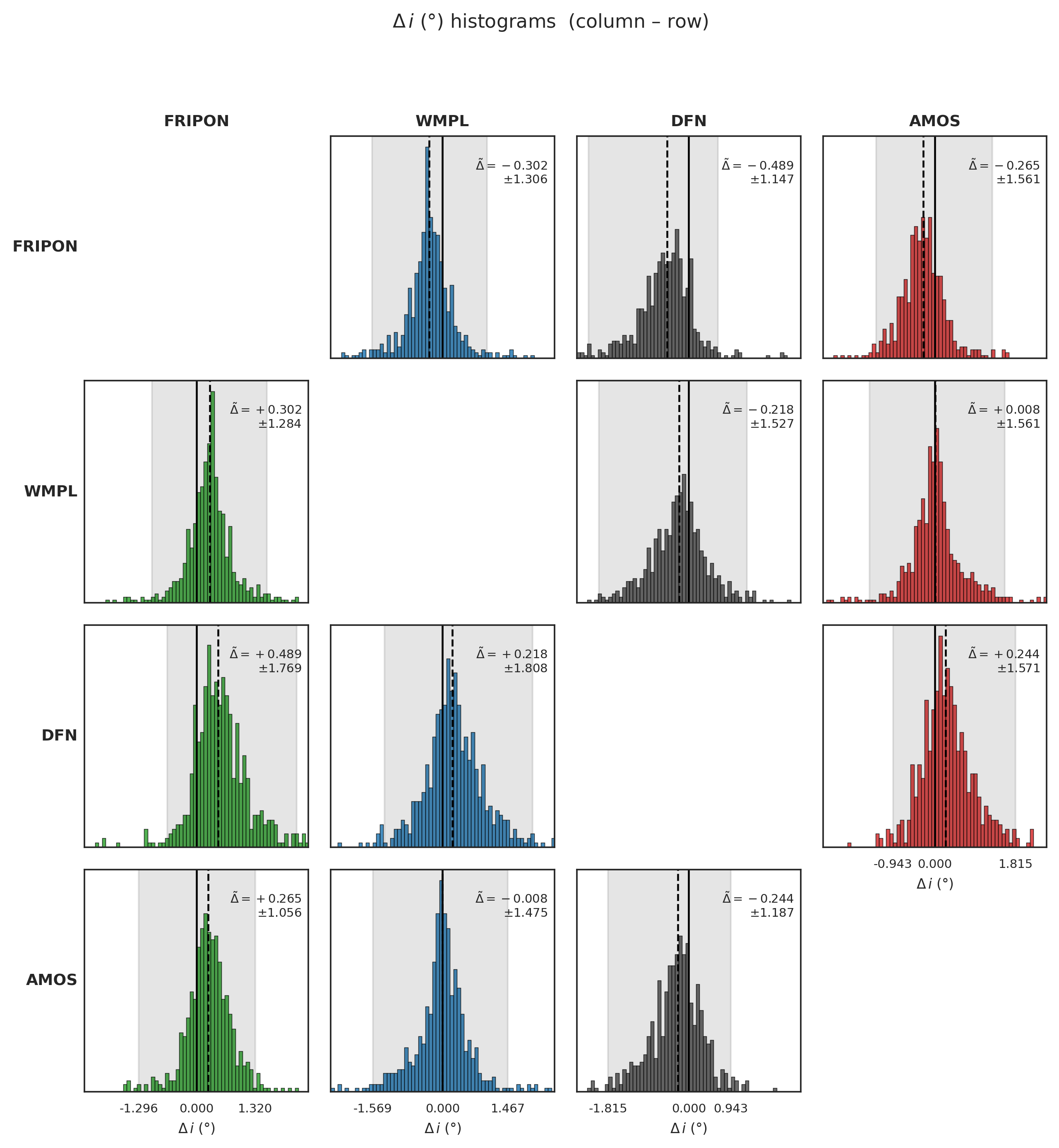}
  \caption{%
    $\Delta i$ histograms (column – row) for all pairwise combinations of FRIPON, WMPL, DFN, and AMOS. The solid vertical line denotes zero, the dashed vertical line indicates the median, and the grey region encompasses the 95\% confidence region. $\Delta$ is the median value, and the $\pm$ indicates half of the 95\% region.
  }
  \label{fig:app_dinc}
\end{figure*}

\begin{figure*}[p]         
  \centering
  \includegraphics[width=\textwidth]{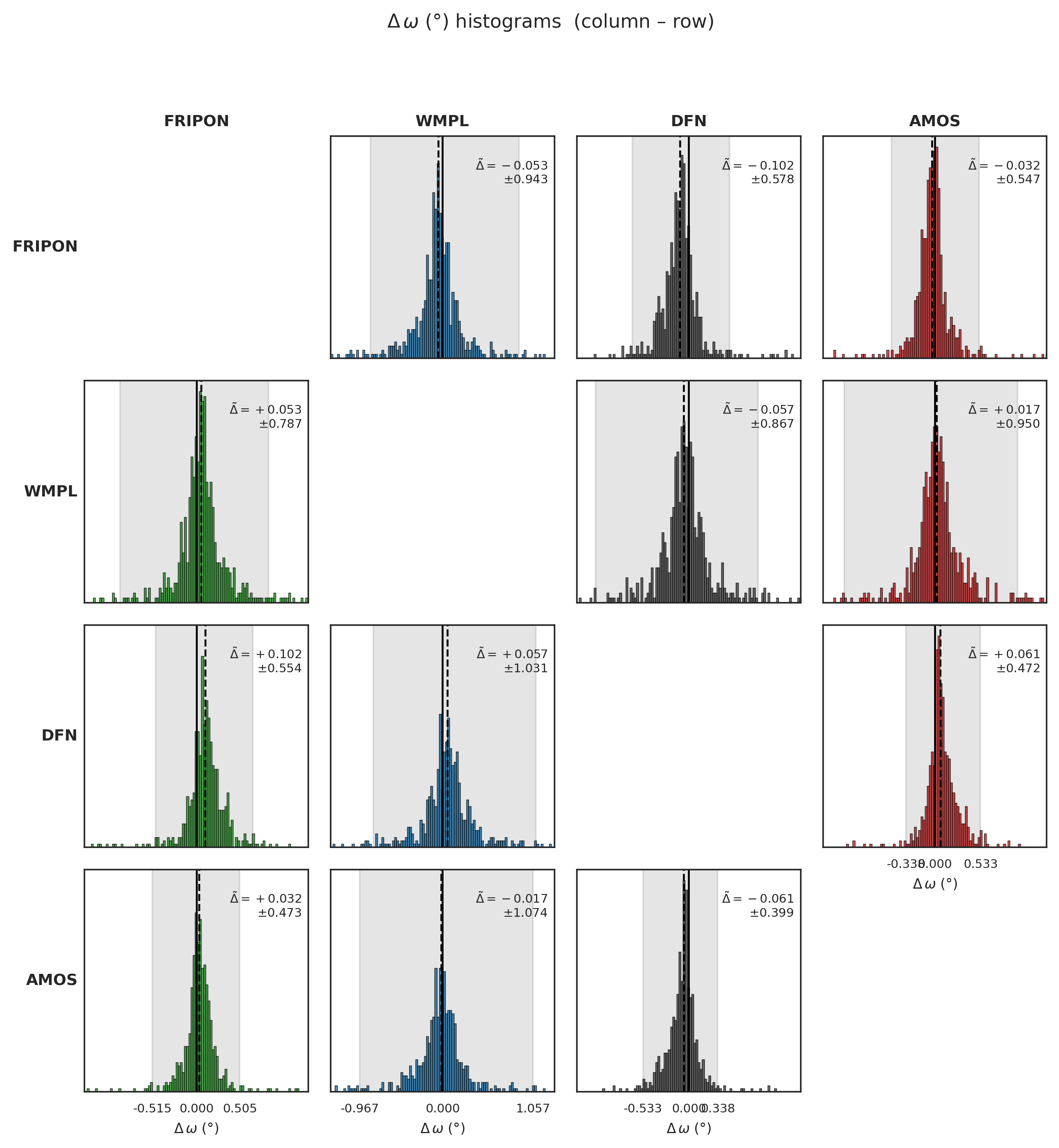}
  \caption{%
    $\Delta \omega$ histograms (column – row) for all pairwise combinations of FRIPON, WMPL, DFN, and AMOS. The solid vertical line denotes zero, the dashed vertical line indicates the median, and the grey region encompasses the 95\% confidence region. $\Delta$ is the median value, and the $\pm$ indicates half of the 95\% region.
  }
  \label{fig:app_domega}
\end{figure*}

\begin{figure*}[p]         
  \centering
  \includegraphics[width=\textwidth]{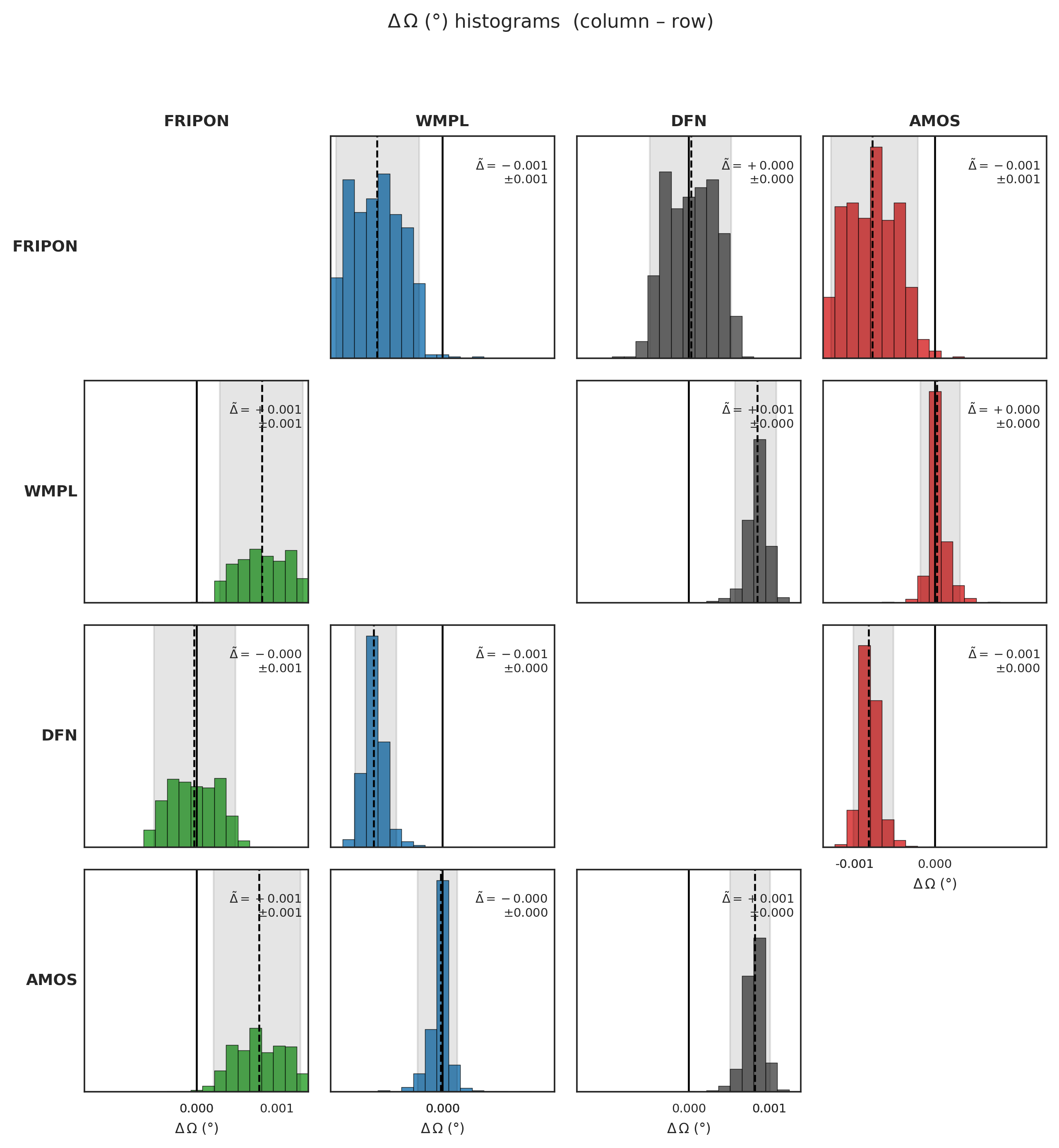}
  \caption{%
    $\Delta \Omega$ histograms (column – row) for all pairwise combinations of FRIPON, WMPL, DFN, and AMOS. The solid vertical line denotes zero, the dashed vertical line indicates the median, and the grey region encompasses the 95\% confidence region. $\Delta$ is the median value, and the $\pm$ indicates half of the 95\% region.
  }
  \label{fig:app_dOmega}
\end{figure*}

\end{appendix}

\end{document}